\documentclass[twocolumn,letterpaper,aps,prc,longbibliography,superscriptaddress,showpacs,nofootinbib,floatfix]{revtex4-1}

\usepackage{graphicx}	
\usepackage{xspace}	

\newcommand{\taa}{\mbox{$T_{AA}$}\xspace}

\newcommand{\pt}{\mbox{$p_{T}$}\xspace}

\newcommand{\raa}{\mbox{$R_{AA}$}\xspace}

\newcommand{\Npart}{\mbox{$N_{\rm part}$}\xspace}
\newcommand{\Ncoll}{\mbox{$N_{\rm coll}$}\xspace}
\newcommand{\Nqp}{\mbox{$N_{\rm qp}$}\xspace}

\newcommand{\dNdeta}{\mbox{$dN_{\rm ch}/d\eta$}\xspace}

\newcommand{\ebjtau}{\mbox{$\varepsilon_{\rm Bj}\tau_{0}$}\xspace}

\newcommand{\sqsn}{\mbox{$\sqrt{s_{_{NN}}}$}\xspace}
\newcommand{\snn}{\mbox{$\sqrt{s_{_{NN}}}$}\xspace}

\newcommand{\pp}{\mbox{$p$$+$$p$}\xspace}
\renewcommand{\AA}{\mbox{$A$$+$$A$}\xspace}

\newcommand{\auau}{\mbox{Au$+$Au}\xspace}
\newcommand{\cucu}{\mbox{Cu$+$Cu}\xspace}
\newcommand{\pbpb}{\mbox{Pb$+$Pb}\xspace}

\newcommand{\sloss}{\mbox{$S_{\rm loss}$}\xspace}
\newcommand{\piz}{\mbox{$\pi^0$}\xspace}
\newcommand{\dptpt}{\mbox{$\delta p_T/p_T$}\xspace}
\newcommand{\dpt}{\mbox{$\delta p_T$}\xspace}
\newcommand{\ptpp}{\mbox{$p_T^{pp}$}\xspace}
\newcommand{\ptaa}{\mbox{$p_T^{AA}$}\xspace}
\newcommand{\gevc}{\mbox{GeV/$c$}\xspace}

\begin{document}

\title{Scaling properties of fractional momentum loss of high-pT hadrons in
nucleus-nucleus collisions at $\sqrt{s_{_{NN}}}$ from 62.4~GeV to 2.76~TeV}

\newcommand{\abilene}{Abilene Christian University, Abilene, Texas 79699, USA}
\newcommand{\augie}{Department of Physics, Augustana University, Sioux Falls, South Dakota 57197, USA}
\newcommand{\banaras}{Department of Physics, Banaras Hindu University, Varanasi 221005, India}
\newcommand{\barc}{Bhabha Atomic Research Centre, Bombay 400 085, India}
\newcommand{\baruch}{Baruch College, City University of New York, New York, New York, 10010 USA}
\newcommand{\bnlcoll}{Collider-Accelerator Department, Brookhaven National Laboratory, Upton, New York 11973-5000, USA}
\newcommand{\bnlphys}{Physics Department, Brookhaven National Laboratory, Upton, New York 11973-5000, USA}
\newcommand{\caucr}{University of California-Riverside, Riverside, California 92521, USA}
\newcommand{\charlesczech}{Charles University, Ovocn\'{y} trh 5, Praha 1, 116 36, Prague, Czech Republic}
\newcommand{\chonbuk}{Chonbuk National University, Jeonju, 561-756, Korea}
\newcommand{\ciae}{Science and Technology on Nuclear Data Laboratory, China Institute of Atomic Energy, Beijing 102413, P.~R.~China}
\newcommand{\cns}{Center for Nuclear Study, Graduate School of Science, University of Tokyo, 7-3-1 Hongo, Bunkyo, Tokyo 113-0033, Japan}
\newcommand{\colorado}{University of Colorado, Boulder, Colorado 80309, USA}
\newcommand{\columbia}{Columbia University, New York, New York 10027 and Nevis Laboratories, Irvington, New York 10533, USA}
\newcommand{\czechtech}{Czech Technical University, Zikova 4, 166 36 Prague 6, Czech Republic}
\newcommand{\dapnia}{Dapnia, CEA Saclay, F-91191, Gif-sur-Yvette, France}
\newcommand{\debrecen}{Debrecen University, H-4010 Debrecen, Egyetem t{\'e}r 1, Hungary}
\newcommand{\elte}{ELTE, E{\"o}tv{\"o}s Lor{\'a}nd University, H-1117 Budapest, P{\'a}zm{\'a}ny P.~s.~1/A, Hungary}
\newcommand{\ewha}{Ewha Womans University, Seoul 120-750, Korea}
\newcommand{\fit}{Florida Institute of Technology, Melbourne, Florida 32901, USA}
\newcommand{\fsu}{Florida State University, Tallahassee, Florida 32306, USA}
\newcommand{\gsu}{Georgia State University, Atlanta, Georgia 30303, USA}
\newcommand{\hanyang}{Hanyang University, Seoul 133-792, Korea}
\newcommand{\hiroshima}{Hiroshima University, Kagamiyama, Higashi-Hiroshima 739-8526, Japan}
\newcommand{\howard}{Department of Physics and Astronomy, Howard University, Washington, DC 20059, USA}
\newcommand{\ihepprot}{IHEP Protvino, State Research Center of Russian Federation, Institute for High Energy Physics, Protvino, 142281, Russia}
\newcommand{\illuiuc}{University of Illinois at Urbana-Champaign, Urbana, Illinois 61801, USA}
\newcommand{\inrras}{Institute for Nuclear Research of the Russian Academy of Sciences, prospekt 60-letiya Oktyabrya 7a, Moscow 117312, Russia}
\newcommand{\instpasczech}{Institute of Physics, Academy of Sciences of the Czech Republic, Na Slovance 2, 182 21 Prague 8, Czech Republic}
\newcommand{\isu}{Iowa State University, Ames, Iowa 50011, USA}
\newcommand{\jaea}{Advanced Science Research Center, Japan Atomic Energy Agency, 2-4 Shirakata Shirane, Tokai-mura, Naka-gun, Ibaraki-ken 319-1195, Japan}
\newcommand{\jinrdubna}{Joint Institute for Nuclear Research, 141980 Dubna, Moscow Region, Russia}
\newcommand{\jyvaskyla}{Helsinki Institute of Physics and University of Jyv{\"a}skyl{\"a}, P.O.Box 35, FI-40014 Jyv{\"a}skyl{\"a}, Finland}
\newcommand{\karoly}{K\'aroly R\'oberts University College, H-3200 Gy\"ngy\"os, M\'atrai\'ut 36, Hungary}
\newcommand{\kek}{KEK, High Energy Accelerator Research Organization, Tsukuba, Ibaraki 305-0801, Japan}
\newcommand{\korea}{Korea University, Seoul, 136-701, Korea}
\newcommand{\kurchatov}{National Research Center ``Kurchatov Institute", Moscow, 123098 Russia}
\newcommand{\kyoto}{Kyoto University, Kyoto 606-8502, Japan}
\newcommand{\labllr}{Laboratoire Leprince-Ringuet, Ecole Polytechnique, CNRS-IN2P3, Route de Saclay, F-91128, Palaiseau, France}
\newcommand{\lahorelums}{Physics Department, Lahore University of Management Sciences, Lahore 54792, Pakistan}
\newcommand{\lawllnl}{Lawrence Livermore National Laboratory, Livermore, California 94550, USA}
\newcommand{\losalamos}{Los Alamos National Laboratory, Los Alamos, New Mexico 87545, USA}
\newcommand{\lpc}{LPC, Universit{\'e} Blaise Pascal, CNRS-IN2P3, Clermont-Fd, 63177 Aubiere Cedex, France}
\newcommand{\lund}{Department of Physics, Lund University, Box 118, SE-221 00 Lund, Sweden}
\newcommand{\maryland}{University of Maryland, College Park, Maryland 20742, USA}
\newcommand{\mass}{Department of Physics, University of Massachusetts, Amherst, Massachusetts 01003-9337, USA}
\newcommand{\michigan}{Department of Physics, University of Michigan, Ann Arbor, Michigan 48109-1040, USA}
\newcommand{\muenster}{Institut f\"ur Kernphysik, University of Muenster, D-48149 Muenster, Germany}
\newcommand{\muhlenberg}{Muhlenberg College, Allentown, Pennsylvania 18104-5586, USA}
\newcommand{\myongji}{Myongji University, Yongin, Kyonggido 449-728, Korea}
\newcommand{\nagasaki}{Nagasaki Institute of Applied Science, Nagasaki-shi, Nagasaki 851-0193, Japan}
\newcommand{\nara}{Nara Women's University, Kita-uoya Nishi-machi Nara 630-8506, Japan}
\newcommand{\natmephi}{National Research Nuclear University, MEPhI, Moscow Engineering Physics Institute, Moscow, 115409, Russia}
\newcommand{\newmex}{University of New Mexico, Albuquerque, New Mexico 87131, USA}
\newcommand{\nmsu}{New Mexico State University, Las Cruces, New Mexico 88003, USA}
\newcommand{\ohio}{Department of Physics and Astronomy, Ohio University, Athens, Ohio 45701, USA}
\newcommand{\ornl}{Oak Ridge National Laboratory, Oak Ridge, Tennessee 37831, USA}
\newcommand{\orsay}{IPN-Orsay, Univ. Paris-Sud, CNRS/IN2P3, Universit\'e Paris-Saclay, BP1, F-91406, Orsay, France}
\newcommand{\peking}{Peking University, Beijing 100871, P.~R.~China}
\newcommand{\pnpi}{PNPI, Petersburg Nuclear Physics Institute, Gatchina, Leningrad region, 188300, Russia}
\newcommand{\riken}{RIKEN Nishina Center for Accelerator-Based Science, Wako, Saitama 351-0198, Japan}
\newcommand{\rikjrbrc}{RIKEN BNL Research Center, Brookhaven National Laboratory, Upton, New York 11973-5000, USA}
\newcommand{\rikkyo}{Physics Department, Rikkyo University, 3-34-1 Nishi-Ikebukuro, Toshima, Tokyo 171-8501, Japan}
\newcommand{\saispbstu}{Saint Petersburg State Polytechnic University, St.~Petersburg, 195251 Russia}
\newcommand{\saopaulo}{Universidade de S{\~a}o Paulo, Instituto de F\'{\i}sica, Caixa Postal 66318, S{\~a}o Paulo CEP05315-970, Brazil}
\newcommand{\seoulnat}{Department of Physics and Astronomy, Seoul National University, Seoul 151-742, Korea}
\newcommand{\stonybrkc}{Chemistry Department, Stony Brook University, SUNY, Stony Brook, New York 11794-3400, USA}
\newcommand{\stonycrkp}{Department of Physics and Astronomy, Stony Brook University, SUNY, Stony Brook, New York 11794-3800, USA}
\newcommand{\subatech}{SUBATECH (Ecole des Mines de Nantes, CNRS-IN2P3, Universit{\'e} de Nantes) BP 20722-44307, Nantes, France}
\newcommand{\tenn}{University of Tennessee, Knoxville, Tennessee 37996, USA}
\newcommand{\titech}{Department of Physics, Tokyo Institute of Technology, Oh-okayama, Meguro, Tokyo 152-8551, Japan}
\newcommand{\tsukuba}{Center for Integrated Research in Fundamental Science and Engineering, University of Tsukuba, Tsukuba, Ibaraki 305, Japan}
\newcommand{\vandy}{Vanderbilt University, Nashville, Tennessee 37235, USA}
\newcommand{\waseda}{Waseda University, Advanced Research Institute for Science and Engineering, 17  Kikui-cho, Shinjuku-ku, Tokyo 162-0044, Japan}
\newcommand{\weizmann}{Weizmann Institute, Rehovot 76100, Israel}
\newcommand{\wigner}{Institute for Particle and Nuclear Physics, Wigner Research Centre for Physics, Hungarian Academy of Sciences (Wigner RCP, RMKI) H-1525 Budapest 114, POBox 49, Budapest, Hungary}
\newcommand{\yonsei}{Yonsei University, IPAP, Seoul 120-749, Korea}
\newcommand{\zagreb}{University of Zagreb, Faculty of Science, Department of Physics, Bijeni\v{c}ka 32, HR-10002 Zagreb, Croatia}
\affiliation{\abilene}
\affiliation{\augie}
\affiliation{\banaras}
\affiliation{\barc}
\affiliation{\baruch}
\affiliation{\bnlcoll}
\affiliation{\bnlphys}
\affiliation{\caucr}
\affiliation{\charlesczech}
\affiliation{\chonbuk}
\affiliation{\ciae}
\affiliation{\cns}
\affiliation{\colorado}
\affiliation{\columbia}
\affiliation{\czechtech}
\affiliation{\dapnia}
\affiliation{\debrecen}
\affiliation{\elte}
\affiliation{\ewha}
\affiliation{\fit}
\affiliation{\fsu}
\affiliation{\gsu}
\affiliation{\hanyang}
\affiliation{\hiroshima}
\affiliation{\howard}
\affiliation{\ihepprot}
\affiliation{\illuiuc}
\affiliation{\inrras}
\affiliation{\instpasczech}
\affiliation{\isu}
\affiliation{\jaea}
\affiliation{\jinrdubna}
\affiliation{\jyvaskyla}
\affiliation{\karoly}
\affiliation{\kek}
\affiliation{\korea}
\affiliation{\kurchatov}
\affiliation{\kyoto}
\affiliation{\labllr}
\affiliation{\lahorelums}
\affiliation{\lawllnl}
\affiliation{\losalamos}
\affiliation{\lpc}
\affiliation{\lund}
\affiliation{\maryland}
\affiliation{\mass}
\affiliation{\michigan}
\affiliation{\muenster}
\affiliation{\muhlenberg}
\affiliation{\myongji}
\affiliation{\nagasaki}
\affiliation{\nara}
\affiliation{\natmephi}
\affiliation{\newmex}
\affiliation{\nmsu}
\affiliation{\ohio}
\affiliation{\ornl}
\affiliation{\orsay}
\affiliation{\peking}
\affiliation{\pnpi}
\affiliation{\riken}
\affiliation{\rikjrbrc}
\affiliation{\rikkyo}
\affiliation{\saispbstu}
\affiliation{\saopaulo}
\affiliation{\seoulnat}
\affiliation{\stonybrkc}
\affiliation{\stonycrkp}
\affiliation{\subatech}
\affiliation{\tenn}
\affiliation{\titech}
\affiliation{\tsukuba}
\affiliation{\vandy}
\affiliation{\waseda}
\affiliation{\weizmann}
\affiliation{\wigner}
\affiliation{\yonsei}
\affiliation{\zagreb}
\author{A.~Adare} \affiliation{\colorado} 
\author{S.~Afanasiev} \affiliation{\jinrdubna} 
\author{C.~Aidala} \affiliation{\columbia} \affiliation{\losalamos} \affiliation{\mass} \affiliation{\michigan} 
\author{N.N.~Ajitanand} \affiliation{\stonybrkc} 
\author{Y.~Akiba} \affiliation{\riken} \affiliation{\rikjrbrc} 
\author{R.~Akimoto} \affiliation{\cns} 
\author{H.~Al-Bataineh} \affiliation{\nmsu} 
\author{J.~Alexander} \affiliation{\stonybrkc} 
\author{M.~Alfred} \affiliation{\howard} 
\author{H.~Al-Ta'ani} \affiliation{\nmsu} 
% ...
\collaboration{PHENIX Collaboration} \noaffiliation

\date{\today}

%------------------------------------------------------------------------------|

\begin{abstract}

%\linenumbers

Measurements of the fractional momentum loss 
($S_{\rm loss}\equiv{\delta}p_T/p_T$) of 
high-transverse-momentum-identified hadrons in heavy ion collisions are 
presented.  Using $\pi^0$ in Au$+$Au and Cu$+$Cu collisions at 
$\sqrt{s_{_{NN}}}=62.4$ and 200~GeV measured by the PHENIX experiment at 
the Relativistic Heavy Ion Collider and and charged hadrons in Pb$+$Pb 
collisions measured by the ALICE experiment at the Large Hadron Collider, 
we studied the scaling properties of $S_{\rm loss}$ as a function of a 
number of variables: the number of participants, $N_{\rm part}$, the 
number of quark participants, $N_{\rm qp}$, the charged-particle density, 
$dN_{\rm ch}/d\eta$, and the Bjorken energy density times the 
equilibration time, $\varepsilon_{\rm Bj}\tau_{0}$.  We find that the 
$p_T$, where $S_{\rm loss}$ has its maximum, varies both with centrality 
and collision energy.  Above the maximum, $S_{\rm loss}$ tends to follow a 
power-law function with all four scaling variables. The data at 
$\sqrt{s_{_{NN}}}=200$~GeV and 2.76~TeV, for sufficiently high particle 
densities, have a common scaling of $S_{\rm loss}$ with $dN_{\rm 
ch}/d\eta$ and $\varepsilon_{\rm Bj}\tau_{0}$, lending insight on the 
physics of parton energy loss.

\end{abstract}

	

\maketitle

		\section{Introduction}

It has been firmly established that in relativistic heavy ion collisions a 
hot, dense medium is rapidly formed, capable of interacting with the high 
\pt partons produced in primordial hard scattering and making them lose 
some energy while traversing the
medium~\cite{Adcox:2004mh, Arsene:2004fa, Back:2004je, Adams:2005dq}.
Such energy loss in the medium was first predicted in early 
1980's~\cite{Bjorken:1982tu}. Quantifying this energy loss is an important 
issue, because it is directly connected to the properties of the medium. 
However, this is not straightforward since neither the original parton 
energy, nor that of the decelerated one is easily accessible.  
Back-to-back photon-jet pairs in principle give access to both the initial 
and final parton energy, but such events are rare, because they are 
suppressed by a factor $\alpha$, the electromagnetic coupling 
constant. Measurement of jets give more complete information on the parton 
energy loss, however, their measurement is challenging, particularly at 
high multiplicities and low parton \pt. To circumvent this, high \pt 
hadrons are often used as proxies for jets (``leading hadrons''), and the 
parton energy loss in principle can be calculated by proper comparison of 
the invariant yields of hadrons in \pp and \AA at a given \pt. For this 
purpose the \pp yields are usually scaled up by the expected number of 
binary nucleon-nucleon collisions in \AA, estimated from a Glauber 
Monte-Carlo model, and in the absence of any initial or final state 
nuclear effects they are expected to coincide with the \AA yields. The 
partons have steeply falling momentum spectra, so if partons lose energy, 
that results in a shift of the momentum spectra, and the yield at a given 
pT will become suppressed~\cite{Wang:1998bha}. Utilizing this fact, the 
nuclear-modification factor (\raa) has become a widely used 
characterization of the energy loss which is defined as:

  \begin{equation}
  R_{\rm AA} (\pt) =  
  \frac{(1/N_{\rm AA}^{\rm evt}) {\rm d}^2N_{\rm AA}^{h}/{\rm d}\pt{\rm dy}}
{\left<T_{\rm AA}\right>\times{\rm d}^2\sigma_{\rm pp}^{h}/{\rm d}\pt{\rm dy
}},
  \end{equation}
\noindent
where $\sigma_{pp}^{h}$ is the production cross section of the respective 
hadron in \pp collisions, $\left<T_{\rm AA}\right> = \left<N_{\rm 
coll}\right>/\sigma_{pp}^{\rm inel}$ is the nuclear overlap function 
averaged over the relevant range of impact parameters, and $\left<N_{\rm 
coll}\right>$ is the number of binary nucleon-nucleon collisions computed 
with $\sigma_{pp}^{\rm inel}$. If \raa is unity, it is usually assumed 
that the yield measured in \AA collisions is explained by the primordial hard 
production as observed in \pp collisions with no nuclear or medium effect. 
If \raa~$<$~1 (suppression) the \AA yield at a given \pt is less than 
that expected from the scaled \pp.

While the parton energy loss is expected to depend both on system size and 
collision energy, it is remarkable that \raa is very similar from 
\sqsn=~62.4 to  200~GeV at the Relativistic Heavy Ion Collider (RHIC)
and up to 2.76~TeV at the Large Hadron Collider (LHC). The 
reason is 
that while the energy loss increases with increasing $\sqrt{s_{NN}}$ which 
would tend to decrease \raa, the power $n$ in the $\pt^{-n}$ shaped 
spectra decreases ($n=10.6$ for 62.4~GeV~\cite{Adare:2012uk}, $n=8.06$ for
200~GeV \auau and $n\approx6.0$ for 2.76~TeV~\cite{Adare:2012wg})
and provides a countervailing effect. 
A numerical calculation showed that the fractional energy loss of 
partons, $\Delta E/E$, is indeed significantly different between LHC and 
RHIC even though the \raa is similar~\cite{Horowitz:2011gd}.

Instead of \raa one can employ the fractional momentum loss (\sloss) of 
high \pt hadrons as a measure of parton energy loss which should reflect 
the average fractional energy loss of the initial partons 
($\left<\Delta E/E\right>\sim\sloss$). \sloss is defined as
\begin{equation}
\sloss 
\equiv \dptpt = \frac{\ptpp - \ptaa}{\ptpp}
\label{Eq2}
\end{equation}
where \ptaa is the \pt of the \AA measurement and \ptpp is that of the \pp 
measurement scaled by the nuclear overlap function \taa of the 
corresponding \AA centrality class at the same yield of the \AA 
measurement. We calculate \sloss as a function of the original momentum of 
partons that are represented by \ptpp.

Under the assumptions that \Ncoll scaling is applicable and fragmentation 
functions are unchanged from \pp collisions, $\delta\pt$ can be directly 
measured as the shift in \pt needed to get the same yield ($dN/dp_{\rm 
T}dy$) in \AA as the scaled \pp.

The PHENIX experiment published a study of the energy loss of partons by 
converting azimuthal angle ($\phi$)-dependent \raa with respect to the 
event plane to \sloss assuming that the spectra follow a power-law 
function~\cite{Adler:2006bw}. That study found that \sloss scales with 
$L_{\epsilon}$, the distance from the center to the edge of the collision 
area which the partons traverse, for all centrality classes for 
3$<\pt<$8~\gevc, and also with the density-weighted path length $\rho 
L/\rho_{\rm cent}$ where $\rho_{\rm cent}$ is the density at the center of the 
collision zone and the $\rho$ is the density at the given coordinate. The 
dependence of \sloss on centrality was also reasonably approximated by 
$\Npart^{2/3}$. A similar study has been performed using \pbpb data 
available at LHC and \auau data from RHIC~\cite{Christiansen:2013hya}. The 
authors found that the scaling in~\cite{Adler:2006bw} does not hold at \pt 
higher than 10~\gevc. Other recent publications tried to obtain 
$\phi$-integrated \sloss without assuming the spectral 
shape~\cite{Adare:2012uk,Adare:2012wg}. It was found that \sloss varies by 
a factor of six from 62.4~GeV \auau to 2.76~TeV \pbpb collisions. 

These studies showed that the fractional momentum loss \sloss has a major 
advantage over \raa, in that it allows for a direct comparison of parton 
energy loss between different colliding systems and energies, because it 
eliminates the bias owing to the \sqsn-variation of the exponent, $n$, in 
the power-law spectra of high \pt particles.

These scaling studies are not a replacement for full 
quantum-chromodynamics calculations of parton energy loss that must 
include different quark and gluon admixtures and their different 
fragmentation functions, initial state effects such as nuclear modified 
parton distribution functions, and potentially modified harmonization 
effects.
That said, since \sloss is merely a new representation of the 
experimental measurements, any such theoretical calculation would need to 
describe the observed scalings at the precision of the uncertainties.

In this paper, we extend the previous studies of $\phi$-integrated \sloss 
by including additional data sets both from RHIC and LHC and by plotting 
the fractional momentum loss against several scaling variables to 
characterize the energy loss mechanism. We average over the event plane 
dependence to simplify the analysis. Section~\ref{AnaSection} describes 
the method of calculating \sloss and introduces the global scaling 
variables. In section~\ref{pTCentDepSloss}, we present values for \sloss 
as a function of centrality for a variety of systems and energies. 
Section~\ref{Highlight} presents the main result of this paper, which is 
the study of the scaling behavior of \sloss. We conclude in 
section~\ref{GrandSummary}.

		\section{Dataset and Analysis\label{AnaSection}}

In this section we describe how fractional momentum loss is calculated and 
define the various scaling variables. A summary of the data is given in 
Table~\ref{DataSetSummary}.  For RHIC energies, data from 
the PHENIX experiment for $\pi^0$ in \auau and \cucu collisions both at 
\snn=~200~GeV and 62.4~GeV were 
used~\cite{Adare:2008qa,Adare:2012wg,Adare:2008ad,Adare:2007dg,Adare:2012uk,Adare:2008qb}, 
while for the LHC, data on charged hadrons and pions in \pbpb collisions, 
both at \snn=~2.76~TeV, measured by the ALICE 
experiment~\cite{Abelev:2012hxa,Abelev:2014laa,Abelev:2014ypa,Abelev:2013ala} 
were used. To calculate the fractional momentum loss, \pp data are also 
needed: RHIC data were taken from~\cite{Adare:2007dg,Adare:2008qb}, while 
LHC data were taken from~\cite{Abelev:2013ala}.

%--------------------------------------------  Table_I
\begin{table}
\caption{Summary of data sets used in this analysis. The 
\sqsn=~62.4 and 200~GeV data are from PHENIX at RHIC and the 
\sqsn=~2.76~TeV data from from ALICE at the LHC.
}
\label{DataSetSummary}
\begin{ruledtabular} \begin{tabular}{cccccc}
System & particle & $\sqrt{s_{NN}}$ & year & \pt range & ref. \\
\hline
Au$+$Au & \piz & 200~GeV & 2004 & 1.0--20~GeV/$c$ & \cite{Adare:2008qa}\\
Au$+$Au & \piz & 200~GeV & 2007 & 5.0--20~GeV/$c$ & \cite{Adare:2012wg}\\
Cu$+$Cu & \piz & 200~GeV & 2005 & 1.0--18~GeV/$c$ & \cite{Adare:2008ad}\\
$p$$+$$p$ & \piz & 200~GeV & 2005 & 0.5--20~GeV/$c$ & \cite{Adare:2007dg}\\
\\
Au$+$Au & \piz & 62.4~GeV & 2010 & 1.0--10~GeV/$c$ & \cite{Adare:2012uk}\\
Cu$+$Cu & \piz & 62.4~GeV & 2005 & 1.0--8.0~GeV/$c$ & \cite{Adare:2008ad}\\
$p$$+$$p$ & \piz & 62.4~GeV & 2006 & 0.5--7.0~GeV/$c$ & \cite{Adare:2008qb}\\
\\
Pb$+$Pb & $h^{+/-}$ & 2.76~TeV & 2010 & 0.2--50~GeV/$c$ & \cite{Abelev:2012hxa}\\
Pb$+$Pb & $\pi^{+/-}$ & 2.76~TeV & 2010-2011 & 2.0--20~GeV/$c$ & \cite{Abelev:2014laa}\\
Pb$+$Pb & \piz & 2.76~TeV & 2010 & 0.5--11~GeV/$c$ & \cite{Abelev:2014ypa}\\
$p$$+$$p$ & $h^{+/-}$ & 2.76~TeV & 2009-2011 & 0.2--50~GeV/$c$ & \cite{Abelev:2013ala}\\
$p$$+$$p$ & $\pi^{+/-}$ & 2.76~TeV & 2010-2011 & 2.0--20~GeV/$c$ & \cite{Abelev:2014laa}\\
$p$$+$$p$ & \piz & 2.76~TeV & 2011 & 0.5--11~GeV/$c$ & \cite{Abelev:2014ypa}\\
\end{tabular} \end{ruledtabular}
\end{table}

	\subsection{Fractional momentum loss}

\begin{figure}[!htb]
  \includegraphics[width=1.0\linewidth]{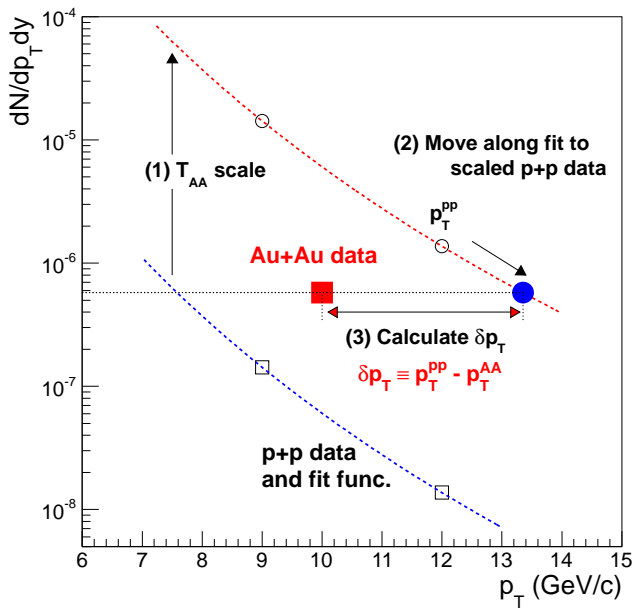}
  \caption{(Color online)
  Method of calculating the fractional momentum loss
  ($S_{{\rm loss}} \equiv$\dptpt).
  This plot is for illustration only; uncertainties are not shown.
  The procedure: (1) scale the \pp data by \taa corresponding
  to the centrality selection of \AA data, 
  (2) fit the \pp data and choose the scaled \pp point closest in yield
  to the \AA along the fit,(3) calculate the difference of scaled \pp
  and \AA transverse momenta, $\dpt\equiv\ptpp-\ptaa$, at the same yield.
  }
    \label{fig:MethoddpTpT}
\end{figure}

Figure~\ref{fig:MethoddpTpT} shows the method of calculating the \sloss 
using measured \AA and $p$$+$$p$ spectra at the same collision energy. First, 
the \piz ($\pi^{+/-}$, $h^{+/-}$) cross section in \pp is scaled by \taa 
corresponding to the centrality selection of the \AA data. Second, the 
scaled \pp cross section is fit with a power-law function. Third, the 
scaled \pp point, \ptpp, corresponding to the yield at the Au$+$Au point 
of interest, is found using the fit to interpolate between scaled \pp 
points.  The $\delta p_T$ is calculated as \ptpp - \ptaa. To obtain 
\sloss, the $\delta p_T$ is divided by \ptpp.

%%%%%%%%%%%%%%%%%%%%%%%%%%%%%%%%%%%% Fig_1

%%%%%%%%%%%%%%%%%%%%%%%%%%%%%%%%%%%% Fig_2
\begin{figure*}[!htb] 
\includegraphics[width=0.8\linewidth]{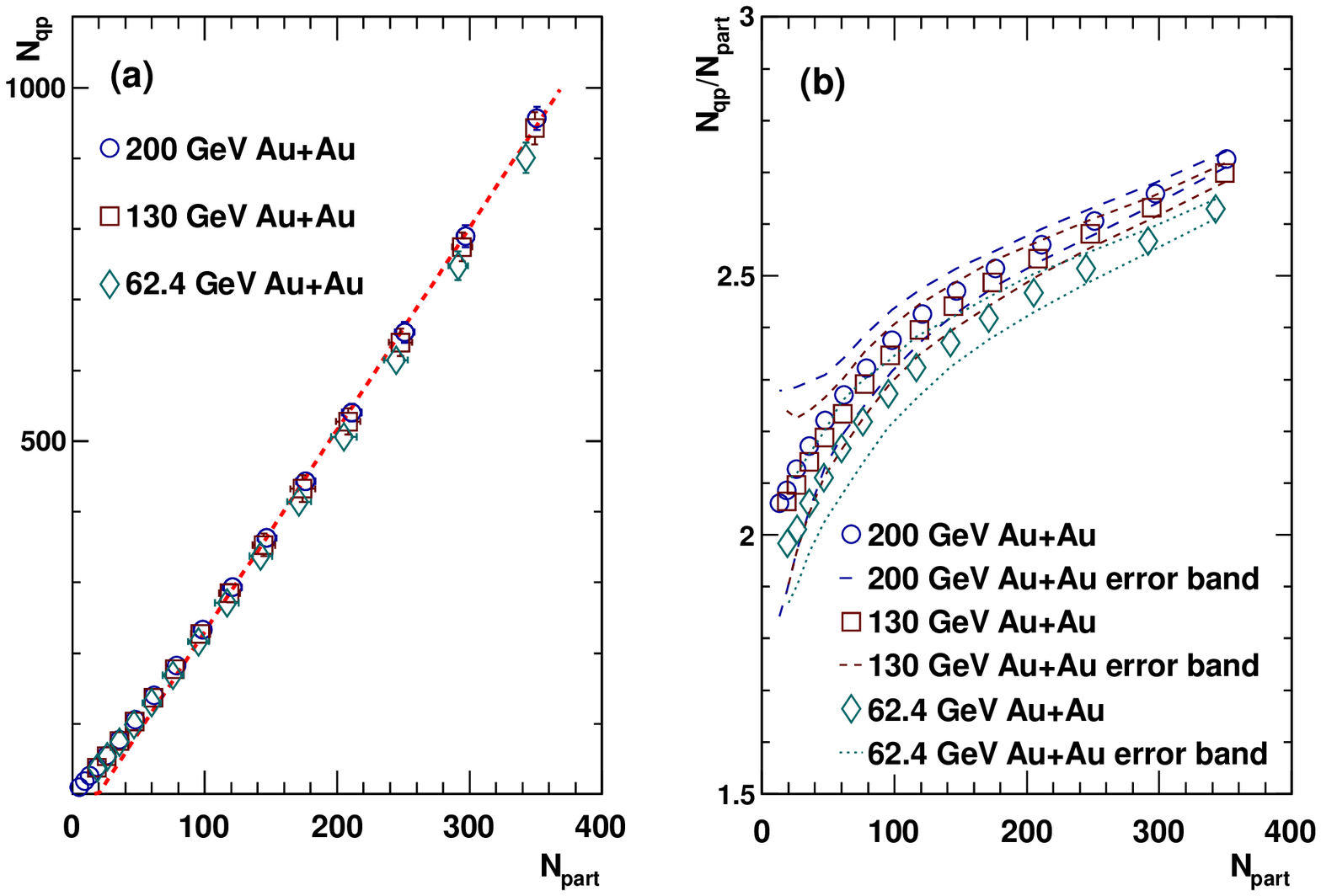}
\caption{(Color online) 
(a) The number of quark participants as a function of the number of 
nucleon participants. The error bars represent the systematic uncertainty 
estimate on the MC-Glauber calculation. The dashed line is a linear fit to 
the 200~GeV Au$+$Au points with $\Npart>100$ to illustrate the 
nonlinearity of the correlation at low values of \Npart. 
(b) The ratio of the number of quark participants to the number of 
nucleon participants as a function of the number of nucleon participants. 
The error bands represent the systematic uncertainty estimate on the 
MC-Glauber calculation.  This figure is reproduced from~\cite{Adler:2013aqf}.
}
    \label{fig:npartVsNq}
\end{figure*}

It is important to realize that the effective fractional energy loss, 
\sloss, estimated from the shift in the \pt spectrum, is actually less 
than the real average energy loss at a given \pt.  This is true because, 
for a given observed \ptaa, the events at much larger \pt with larger 
energy loss are lost under the events at smaller \pt with a 
correspondingly smaller energy loss owing to the steeply falling 
spectrum.  We evaluated this bias to the \sloss measurement with a simple 
Monte Carlo calculation using the power of the spectra obtained in the 
measurements, and found that it is $\sim$10\% for collisions at 
\sqsn=~200~GeV and 62.4~GeV, and $\sim$18\% for \sqsn=~2.76~TeV.  This 
systematic effect is not reflected in the final data uncertainties.

The uncertainties of the \sloss are obtained as follows. We first 
estimated the errors of yields for the \AA and the \pp points in three 
categories; the quadratic sum of the statistical and \pt-independent 
systematic uncertainties (``Type A''), \pt-correlated systematic 
uncertainties (``Type B''), and the overall scale uncertainties which 
allow all the data points to move to the same direction with a certain 
fraction of the central values (``Type C''). The Type B is the quadratic 
sum of the systematic uncertainties related to the measurement of \piz for 
the PHENIX result, including those of photon identification efficiency, 
energy scale, and background subtraction. The Type C is the quadratic sum 
of the \taa and \pp normalization uncertainties in this analysis. The 
uncertainties for the \AA and \pp points in three categories are 
separately summed in quadrature, and projected to the \ptpp axis using the 
\pp fit function.

	\subsection{Number of Nucleon and Quark Participants}

To study the systematics of fractional momentum loss, we 
introduce several scaling variables. Here we briefly describe how the 
number of nucleon participants (\Npart) and quark participants 
(\Nqp)~\cite{Adler:2013aqf} are obtained. The \Npart for the Pb$+$Pb 
collisions at \snn=~2.76~TeV was taken from \cite{Aamodt:2010cz}. The 
number of quark-participants is calculated for all systems as part of this 
work, as explained below.

A Monte-Carlo-Glauber (MC-Glauber) model calculation~\cite{Miller:2007ri} 
is used to obtain estimates for the number of nucleon participants at each 
centrality using the procedure described in~\cite{Adler:2004zn}. A similar 
procedure can be used to estimate the number of quark participants, \Nqp, 
at each centrality~\cite{Adler:2013aqf}. The MC-Glauber calculation is 
modified such that the fundamental interactions are quark-quark rather 
than nucleon-nucleon collisions. The nuclei are assembled by distributing 
the centers of the nucleons according to a Woods-Saxon distribution. Once 
a nucleus is assembled, three quarks are then distributed around the 
center of each nucleon. In our model, we assume the spatial distribution 
of the quarks follows an exponential charge distribution as measured in 
electron-proton elastic scattering:
\begin{equation}
   \rho^{proton}(r) = \rho^{proton}_{0} \times e^{-ar},
\end{equation}
where $a = \sqrt{12}/r_{m} = 4.27$ fm$^{-1}$ and $r_{m}=0.81$ fm is the 
rms charge radius of the proton~\cite{Hofstadter:1956qs}. The coordinates 
of the two colliding nuclei are shifted at random relative to each other 
by a vector $\vec{b}$, the impact parameter, which covers an area larger 
than the maximum possible impact parameter. A pair of quarks, one from 
each nucleus, interact with each other if their distance $d$ in the plane 
transverse to the beam axis satisfies the condition
\begin{equation}
   d < \sqrt{\frac{\sigma^{\rm inel}_{qq}}{\pi}},
\end{equation}

where $\sigma^{\rm inel}_{qq}$ is the inelastic quark-quark cross section, 
which is varied for the case of nucleon-nucleon collisions until the known 
inelastic nucleon-nucleon cross section is reproduced; this
$\sigma^{\rm inel}_{qq}$ is then used for the A+A calculations. 
The inelastic quark-quark cross sections 
are tabulated in Table~\ref{tab:qqCross}. Figure~\ref{fig:npartVsNq}a 
shows the number of quark participants as a function of the number of 
nucleon participants~\cite{Adler:2013aqf}.
The relationship is nonlinear, especially for low values of \Npart. 
The nonlinearity is clearly seen in Fig.~\ref{fig:npartVsNq}b where
the ratio of the number of quark participants to the number of nucleon 
participants as a function of the number of nucleon participants is shown.

%--------------------------------------------  Table_II
\begin{table}[!htb]
\caption{
The inelastic quark-quark cross sections used for each collision energy to 
reproduce the inelastic nucleon-nucleon cross section.}
\label{tab:qqCross}
\begin{ruledtabular}
\begin{ruledtabular} \begin{tabular}{ccccc}
& \sqsn (GeV) & $\sigma^{\rm inel}_{NN}$ (mb) & $\sigma^{\rm inel}_{qq}$ (mb) & \\
\hline
& 2760 & 64.0 & 18.4 & \\
& 200 & 42.3 & 9.36 & \\
& 62.4 & 36.0 & 7.08 & \\
\end{tabular} \end{ruledtabular}
\end{ruledtabular}
\end{table}

	\subsection{Charged Particle Multiplicity}

Another scaling variable used is charged particle multiplicity, or 
multiplicity density, \dNdeta, measured at midrapidity ($y\approx\eta 
\approx 0$).  This quantity is closely related to the gluon density, 
$dN_{\rm gluon}/dy$~\cite{Luzum:2008ch}, as well as to the number of 
participating nucleons \Npart, which in turn is a measure of the system 
size. In a previous publication~\cite{Adler:2004zn} it has been shown 
that \begin{equation} dN_{\rm ch}/d\eta \propto \Npart^\alpha 
\end{equation} where $\alpha$=1.16 in Au$+$Au collisions at \snn=~200~GeV. 
For the RHIC data \dNdeta values were taken from the PHENIX 
experiment~\cite{Adler:2004zn,Adler:2013aqf}, where charged particle 
multiplicities are measured in the $|\eta|<0.35$ pseudorapidity region in 
two pad chamber detectors~\cite{Adcox:2003ct} in zero magnetic 
field.  For the LHC data \dNdeta, values are quoted from the ALICE 
publication~\cite{Aamodt:2010cz}, where charged particles are measured in 
their silicon-pixel detector and quoted in the restricted $|\eta|<0.5$ 
pseudorapidity range.

	\subsection{Bjorken Energy Density}

Finally, we introduce a measure of the energy density. In relativistic 
heavy ion collisions, the Bjorken energy density is frequently used for 
this purpose~\cite{Bjorken:1982qr}. The Bjorken energy density is defined 
as
\begin{equation}
\epsilon_{Bj} = \frac{1}{\tau_0 A_{\perp}} \frac{dE_T}{dy}
\end{equation}
where $\tau_0$ is the proper time when the QGP is equilibrated, 
$A_{\perp}$ is the transverse area of the system. The $A_{\perp}$ can be 
written as $\sim \sigma_x\sigma_y$, where $\sigma_x$ and $\sigma_y$ are 
the widths of $x$ and $y$ position distributions of the participating 
nucleons in the transverse plane, and was estimated using a Monte-Carlo 
Glauber simulation~\cite{Miller:2007ri}. The equilibration time $\tau_0$ 
is strongly model-dependent, therefore, we decided to use \ebjtau as a 
scaling variable, which then contains only well-established experimental 
quantities. The measured $dE_T/d\eta$ is converted to $dE_T/dy$ by 
applying a factor that compensates the phase space difference between 
rapidity and pseudorapidity which is obtained by a simple numerical 
calculation. The factor is found to be 1.25 for \snn=~62.4~GeV and 
\snn=~200~GeV~\cite{Adler:2004zn}, and 1.09 for 
\snn=~2.76~TeV~\cite{Chatrchyan:2012mb}. The uncertainties on these scale 
numbers are $\sim$3\%.  The $dE_T/d\eta$ for the \snn=~2.76~TeV Pb$+$Pb 
collisions are obtained from the literature~\cite{Loizides:2011ys}.

		\section{Results and Discussion\label{ResultAndDiscuss}}

The numerical values of the scaling variables defined in the previous  
section are listed in Table~\ref{GlobalParameters}. 

%--------------------------------------------  Table_III

\begin{table*}[!htb]
\caption{Global variables for Au$+$Au and Cu$+$Cu collisions at 
RHIC from 
PHENIX~\cite{Adare:2008qa,Adare:2008ad,Adare:2012uk,Adare:2012wg}
and Pb$+$Pb collisions at the LHC from
ALICE~\cite{Aamodt:2010jd,Abelev:2012hxa,Abelev:2014laa}.
}
\label{GlobalParameters}
\begin{ruledtabular} \begin{tabular}{ccccccc}
Collision & \sqsn &
Centrality & \Npart & \Nqp & \dNdeta & \ebjtau [GeV/fm$^2$] \\
\hline
Au$+$Au & 200 GeV &
   0\%--5\% & 353$\pm$10.0 & 957$\pm$16.2 & 687$\pm$37.0 & 5.42$\pm$0.59 \\
&& 0\%--10\% & 327$\pm$9.5 & 873$\pm$15.8 & 624$\pm$32.4 & 5.17$\pm$0.56 \\
&& 10\%--20\% & 235$\pm$7.7 & 597$\pm$13.4 & 415$\pm$20.0 & 4.28$\pm$0.47 \\
&& 20\%--30\% & 166$\pm$6.3 & 403$\pm$11.3 & 274$\pm$15.1 & 3.48$\pm$0.40 \\
&& 30\%--40\% & 114$\pm$5.3 & 263$\pm$10.1 & 177$\pm$11.6 & 2.74$\pm$0.34 \\
&& 40\%--50\% & 75.0$\pm$4.5 & 162$\pm$6.1 & 110$\pm$9.2 & 2.06$\pm$0.28 \\
&& 50\%--60\% & 46.4$\pm$4.0 & 91.5$\pm$6.2 & 61.6$\pm$7.1 & 1.38$\pm$0.23 \\
&& 60\%--70\% & 26.1$\pm$3.5 & 51.3$\pm$6.9 & 31.6$\pm$5.0 & 0.83$\pm$0.18 \\
\\
Cu$+$Cu & 200 GeV &
   0\%--10\% & 96.9$\pm$3.9 & 238$\pm$12.2 & 178$\pm$14.2 & 3.00$\pm$0.36 \\
&& 10\%--20\% & 74.3$\pm$3.9 & 175$\pm$10.5 & 123$\pm$9.9 & 2.43$\pm$0.27 \\
&& 20\%--30\% & 53.7$\pm$2.7 & 121$\pm$8.7 & 85.0$\pm$6.8 & 2.00$\pm$0.25 \\
&& 30\%--40\% & 39.9$\pm$3.8 & 87.1$\pm$9.0 & 57.7$\pm$4.6 & 1.58$\pm$0.19 \\
&& 40\%--50\% & 28.1$\pm$3.3 & 59.0$\pm$7.9 & 38.2$\pm$3.0 & 1.24$\pm$0.17 \\
\\
Au$+$Au & 62.4 GeV &
   0\%--10\% & 317$\pm$6.1 & 824$\pm$21.0 & 405$\pm$32.4 & 3.41$\pm$0.36 \\
&& 10\%--20\% & 225$\pm$9.3 & 560$\pm$17.4 & 273$\pm$20.9 & 2.95$\pm$0.30 \\
&& 20\%--40\% & 131$\pm$8.5 & 310$\pm$12.9 & 151$\pm$13.1 & 2.17$\pm$0.22 \\
&& 40\%--60\% & 54.7$\pm$6.0 & 118$\pm$8.0 & 57.5$\pm$4.3 & 1.31$\pm$0.13 \\
\\
Cu$+$Cu & 62.4 GeV &
   0\%--10\% & 95.9$\pm$2.1 & 222$\pm$9.1 & 122$\pm$8.9 & 1.98$\pm$0.22 \\
&& 10\%--20\% & 73.7$\pm$2.6 & 164$\pm$8.4 & 84.5$\pm$6.5 & 1.65$\pm$0.19 \\
&& 20\%--30\% & 55.2$\pm$2.5 & 118$\pm$7.0 & 58.0$\pm$4.5 & 1.35$\pm$0.16 \\
&& 30\%--40\% & 40.5$\pm$2.4 & 83.6$\pm$6.7 & 39.0$\pm$3.0 & 1.10$\pm$0.13 \\
&& 40\%--50\% & 28.2$\pm$2.2 & 56.0$\pm$5.1 & 25.5$\pm$2.0 & 0.89$\pm$0.11 \\
\\
Pb$+$Pb & 2.76 TeV &
   0\%--5\% & 383$\pm$3.1 & 1086$\pm$14.1 & 1601$\pm$60 & 11.5$\pm$1.43 \\
&& 5\%--10\% & 330$\pm$4.6 & 915$\pm$11.9 & 1294$\pm$49 & 10.5$\pm$1.27 \\
&& 10\%--20\% & 261$\pm$4.4 & 706$\pm$10.6 & 966$\pm$37 & 9.05$\pm$1.41 \\
&& 20\%--30\% & 186$\pm$3.9 & 488$\pm$8.3 & 649$\pm$23 & 7.35$\pm$1.21 \\
&& 30\%--40\% & 129$\pm$3.3 & 325$\pm$7.5 & 426$\pm$15 & 5.99$\pm$0.91 \\
&& 40\%--50\% & 85.0$\pm$2.6 & 205$\pm$5.9 & 261$\pm$9 & 4.69$\pm$0.75 \\
&& 50\%--60\% & 52.8$\pm$2.0 & 118$\pm$3.5 & 149$\pm$6 & 3.47$\pm$0.49 \\
&& 60\%--70\% & 30.0$\pm$1.3 & 60.9$\pm$2.0 & 76$\pm$4 & 2.11$\pm$0.35 \\
&& 70\%--80\% & 15.8$\pm$0.6 & 26.3$\pm$0.9 & 35$\pm$2 & 1.17$\pm$0.22 \\
\end{tabular} \end{ruledtabular}
\end{table*}

%%%%%%%%%%%%%%%%%%%%%%%%%%%%%%%%%%%% Fig_3
\begin{figure}[!htb]
  \includegraphics[width=1.0\linewidth]{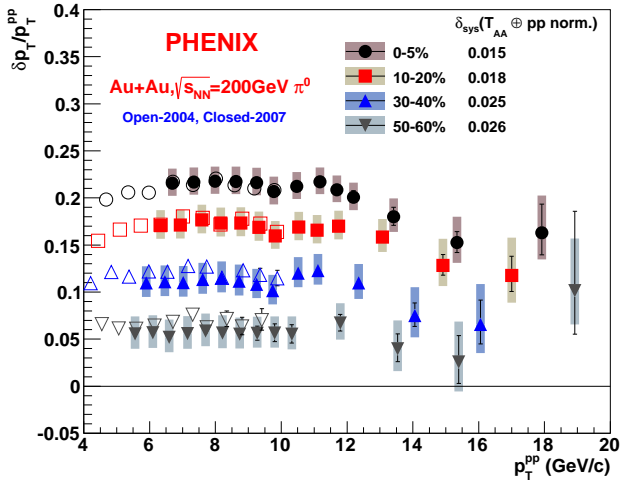}
\caption{(Color online) 
\ptpp dependence of \sloss for $\pi^0$ in 200~GeV Au$+$Au collisions 
from (solid symbols) 2007 data~\protect\cite{Adare:2012wg} and (open 
symbols) 2004 data from the PHENIX experiment at RHIC for 
\pt$<$10~\gevc~\protect\cite{Adare:2008qa}. The error boxes 
corresponding to Type-B errors are not shown for Year-2004 data, but 
the magnitude are same as the ones for Year-2007 data. $\delta_{\rm 
sys}$(\taa$\oplus$~pp norm) are Type-C errors and show the absolute 
amount that the data points would move.
  }
    \label{fig:dpTpTAuAu200eSplit}
\end{figure}

	\subsection{\pt dependence of the fractional momentum loss\label{pTCentDepSloss}}

Figure~\ref{fig:dpTpTAuAu200eSplit} shows the \pt dependence of the 
fractional momentum loss of \piz for various centralities in \auau 
200~GeV collisions, using 2007 data~\cite{Adare:2012wg}. The error bars 
represent the projection of Type A uncertainties to the $p_T^{pp}$ 
axis, while the boxes are the same projection of Type B uncertainties. 
$\delta_{\rm sys}$(\taa$\oplus$~pp norm) shown in the following plots 
stands for the projection of Type C uncertainties to the $p_T^{pp}$ 
axis. Note that $\delta_{\rm sys}$(\taa$\oplus$~pp norm) indicate the 
absolute amount that the data points would move.

The 2007 data set has been analyzed only above \pt=~5~\gevc, which also 
limits the \pt where \sloss can be extracted.  For lower \pt the 2004 
data were used~\cite{Adare:2008qa}, and the results are shown in open 
symbols in Fig.~\ref{fig:dpTpTAuAu200eSplit}. The consistency of 
\raa from 2004 and 2007 data has already been shown in Fig. 11 
of~\cite{Adare:2008qa}. The same consistency can be seen in the 
extracted \sloss. In the central collisions \sloss is slightly 
increasing up to $\sim$6~\gevc, then flattens out and finally 
decreases at the highest measured \pt. As expected, \sloss increases 
monotonically with centrality.

\begin{figure}[!htb]
  \includegraphics[width=1.0\linewidth]{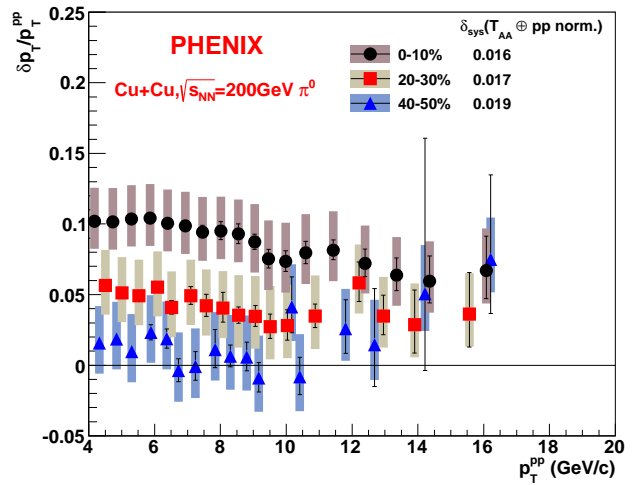}
  \caption{(Color online) 
\ptpp dependence of \sloss for $\pi^0$ in 200~GeV \cucu collisions using 
the spectra measured by PHENIX at RHIC in 2005~\cite{Adare:2008ad}. 
$\delta_{\rm sys}$(\taa$\oplus$~pp norm) are Type-C errors and show the 
absolute amount that the data points would move.
  }
    \label{fig:dpTpTCuCu200e}
\end{figure}

We show the fractional momentum loss of $\pi^0$ for various centralities 
in \cucu 200~GeV collisions in Fig.~\ref{fig:dpTpTCuCu200e}.

%%%%%%%%%%%%%%%%%%%%%%%%%%%%%%%%%%%% Fig_4
We already found in a previous publication that $R_{\rm AA}$ is similar 
at the same \Npart between Cu+Cu and Au$+$Au collisions at 
\sqsn=~200~GeV~\cite{Adare:2008ad}. The \Npart for 0\%--10\% centrality in 
Cu+Cu collisions is similar to the one for 30\%--40\% centrality in Au$+$Au 
collisions. We can see that the \sloss is similar in these collision from 
Figs.~\ref{fig:dpTpTAuAu200eSplit} and ~\ref{fig:dpTpTCuCu200e}.

\begin{figure}[!htb]
  \includegraphics[width=1.0\linewidth]{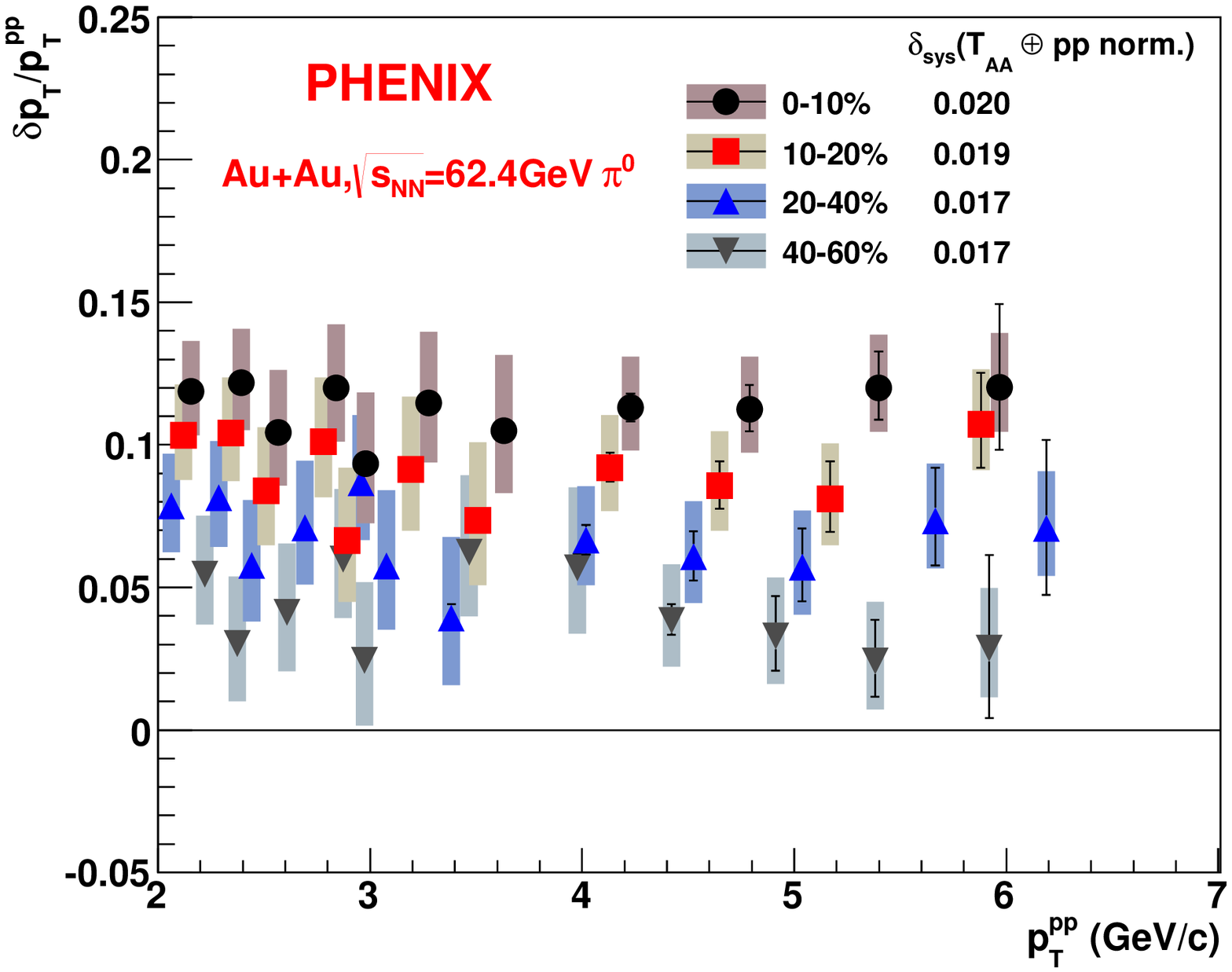}
\caption{(Color online) 
\ptpp dependence of \sloss for $\pi^0$ in 62~GeV \auau collisions using 
the spectra measured by PHENIX in 2010~\cite{Adare:2012uk}. 
$\delta_{\rm 
sys}$(\taa$\oplus$~pp norm) are Type-C errors and show the absolute 
amount that the data points would move.
  }
    \label{fig:dpTpTAuAu62e}
\end{figure}

The fraction of hard-scattering is smaller and therefore results in a 
steeper \pt spectrum at \snn=~62.4~GeV.  Figure~\ref{fig:dpTpTAuAu62e} 
shows the fractional momentum loss of $\pi^0$ for various centralities in 
Au$+$Au 62.4~GeV collisions.

%%%%%%%%%%%%%%%%%%%%%%%%%%%%%%%%%%%% Fig_5

\begin{figure}[!htb]
  \includegraphics[width=1.0\linewidth]{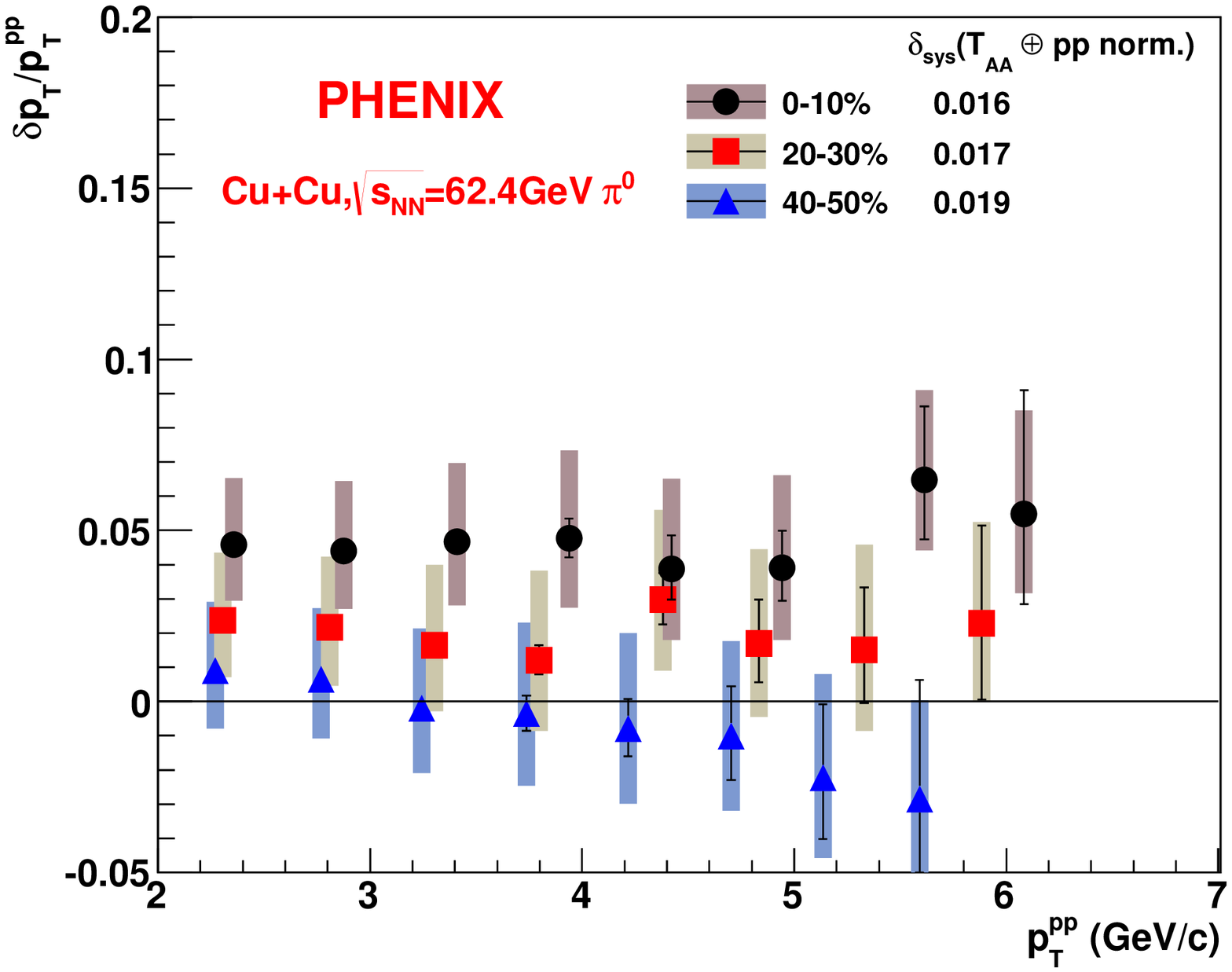}
\caption{(Color online) 
\ptpp dependence of \sloss for $\pi^0$ in 62.4~GeV \cucu collisions using 
the spectra measured by PHENIX in 2005~\cite{Adare:2008ad}. 
$\delta_{\rm sys}$(\taa$\oplus$~pp norm) are Type-C errors and show the 
absolute amount that the data points would move.
  }
    \label{fig:dpTpTCuCu62e}
\end{figure}

The \sloss is much smaller than at 200~GeV even for the most central 
collisions. Note that soft production in $A+A$ collisions still 
contributes to the \ptpp 
range of 2-6~\gevc, where \raa is not reaching to its 
minimum~\cite{Adare:2012uk}. In the \sloss, this will result in smaller 
values. Figure~\ref{fig:dpTpTCuCu62e} shows the \sloss of \piz for various 
centralities in 62.4~GeV \cucu collisions~\cite{Adare:2012uk}.

%%%%%%%%%%%%%%%%%%%%%%%%%%%%%%%%%%%% Fig_6

The trends are similar for the \cucu and \auau collision data. Note that 
in the 62.4~GeV data set the systematic uncertainties from \piz 
reconstruction, overall energy scale and trigger efficiency were 
larger~\cite{Adare:2008ad} than in the 200~GeV \auau data, which explains 
the larger overall systematic uncertainties. It is again interesting to 
mention that within the uncertainties, the 0\%--10\% \cucu collisions give 
the similar \sloss as the 20\%--40\% \auau collisions even at this energy.

\begin{figure}[!htb]
  \includegraphics[width=1.0\linewidth]{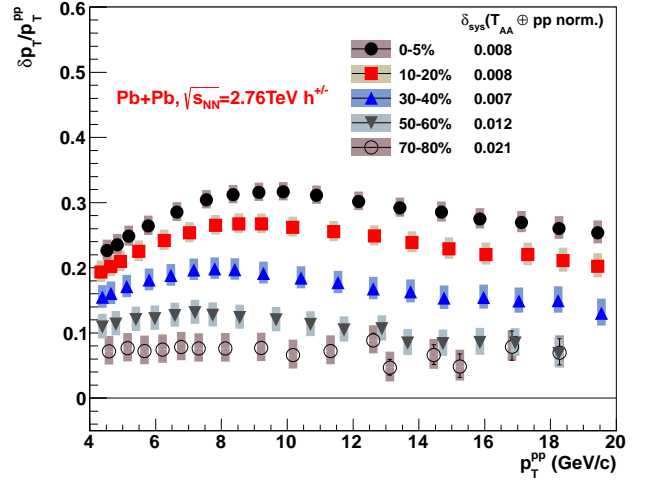}
\caption{(Color online) 
\ptpp dependence of \sloss for charged hadrons in 2.76~TeV \pbpb 
collisions using the result from the ALICE 
experiment~\cite{Abelev:2012hxa,Abelev:2013ala}. $\delta_{\rm 
sys}$(\taa$\oplus$~pp norm) are Type-C errors and show the absolute 
amount that the data points would move.
  }
    \label{fig:dpTpTPbPb276eSplit}
\end{figure}

In Fig.~\ref{fig:dpTpTPbPb276eSplit}, we show the fractional momentum 
loss for charged hadrons in \pbpb collisions at \sqsn = 2.76~TeV measured 
by the ALICE experiment~\cite{Abelev:2012hxa,Abelev:2013ala}.

%%%%%%%%%%%%%%%%%%%%%%%%%%%%%%%%%%%% Fig_7

A clear increase of the \sloss is seen in the 4-10~\gevc region with the 
maximum being dependent on centrality. Despite the $\approx$10\%--fold 
difference of \sqsn between RHIC and LHC, the trend is rather consistent, 
but more pronounced at the LHC and without a region of constant \sloss as 
is most evident in the PHENIX 0\%--10\% data in 
Fig.~\ref{fig:dpTpTAuAu200eSplit}.

%%%%%%%%%%%%%%%%%%%%%%%%%%%%%%%%%%%% Fig_8
\begin{figure}[!htb]
  \includegraphics[width=1.0\linewidth]{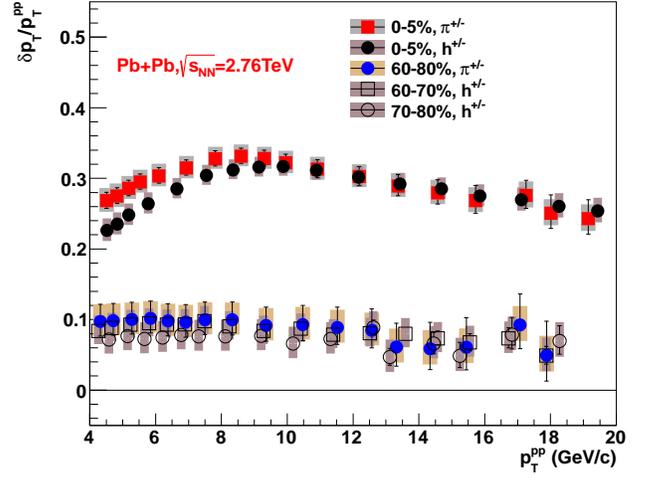}
\caption{(Color online)
\ptpp dependence of \sloss for charged pions in 2.76~TeV \pbpb collisions 
together with those for charged hadrons from the same collision system. 
The charged pion result is from the ALICE 
experiment~\cite{Abelev:2014laa}.
  }
    \label{fig:dpTpTPbPb276ePi}
\end{figure}

\begin{figure}[!htb]
  \includegraphics[width=1.0\linewidth]{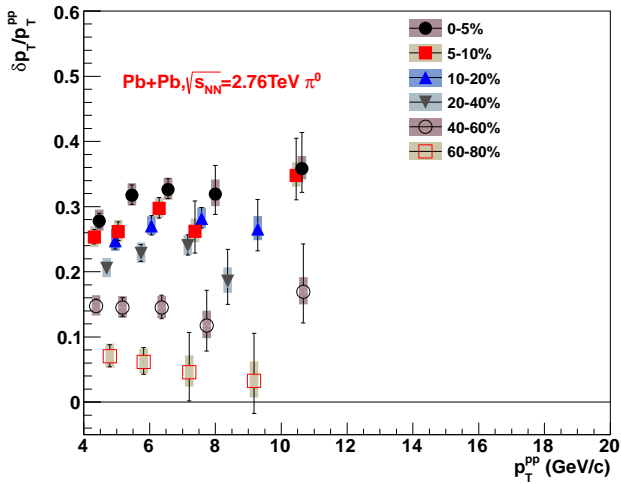}
          \caption{(Color online)\ptpp dependence of \sloss 
	  for neutral pions in 2.76~TeV Pb$+$Pb collisions using
the result from the ALICE experiment~\cite{Abelev:2014ypa}. 
  }
    \label{fig:dpTpTPbPb276ePi0}
\end{figure}

The ALICE experiment recently published the spectra for charged pions for 
two centrality classes~\cite{Abelev:2014laa}. We computed the fractional 
momentum loss for charged pions and compared with those for charged 
hadrons as shown in Fig.~\ref{fig:dpTpTPbPb276ePi}. For peripheral 
collisions, we plot the results for charged hadrons in 60\%--70\% and 
70\%--80\% bins. For 0\%--5~\% centrality, the \sloss for charged hadrons are 
systematically lower than that of charged pions at \pt$<$10~\gevc, and 
both of them become similar above 10~\gevc. This observation is 
consistent with the enhanced baryon production in \pt$<$10~\gevc compared 
to mesons in the central collisions~\cite{Abelev:2014laa}. Charged hadron 
spectra include protons, and thus the suppression is smaller for them in 
the medium \pt region. In the 60\%--80\% centrality, the charged pions and 
charged hadrons give similar results. This feature is again consistent 
with the observation of enhanced baryon production both at RHIC and LHC 
which only occurs in the central collisions. The ALICE experiment also 
published neutral pion data very recently, from which we calculated the 
\sloss for the data set as shown in 
Figure~\ref{fig:dpTpTPbPb276ePi0}~\cite{Abelev:2014ypa}.

%%%%%%%%%%%%%%%%%%%%%%%%%%%%%%%%%%%% Fig_9

The neutral pion results have finer centrality selections, but have a 
limited \pt range and larger uncertainties, therefore, they were not 
considered in further studies of scaling variable dependence. We can see 
that the \sloss for neutral pions are similar to that of charged pions and 
hence are consistent with charged hadrons for \pt$>$10~\gevc.

	\subsection{Scaling variable dependence\label{Highlight}}

%%%%%%%%%%%%%%%%%%%%%%%%%%%%%%%%%%%% Fig_10
\begin{figure*}[!htb]
  \includegraphics[width=0.78\linewidth]{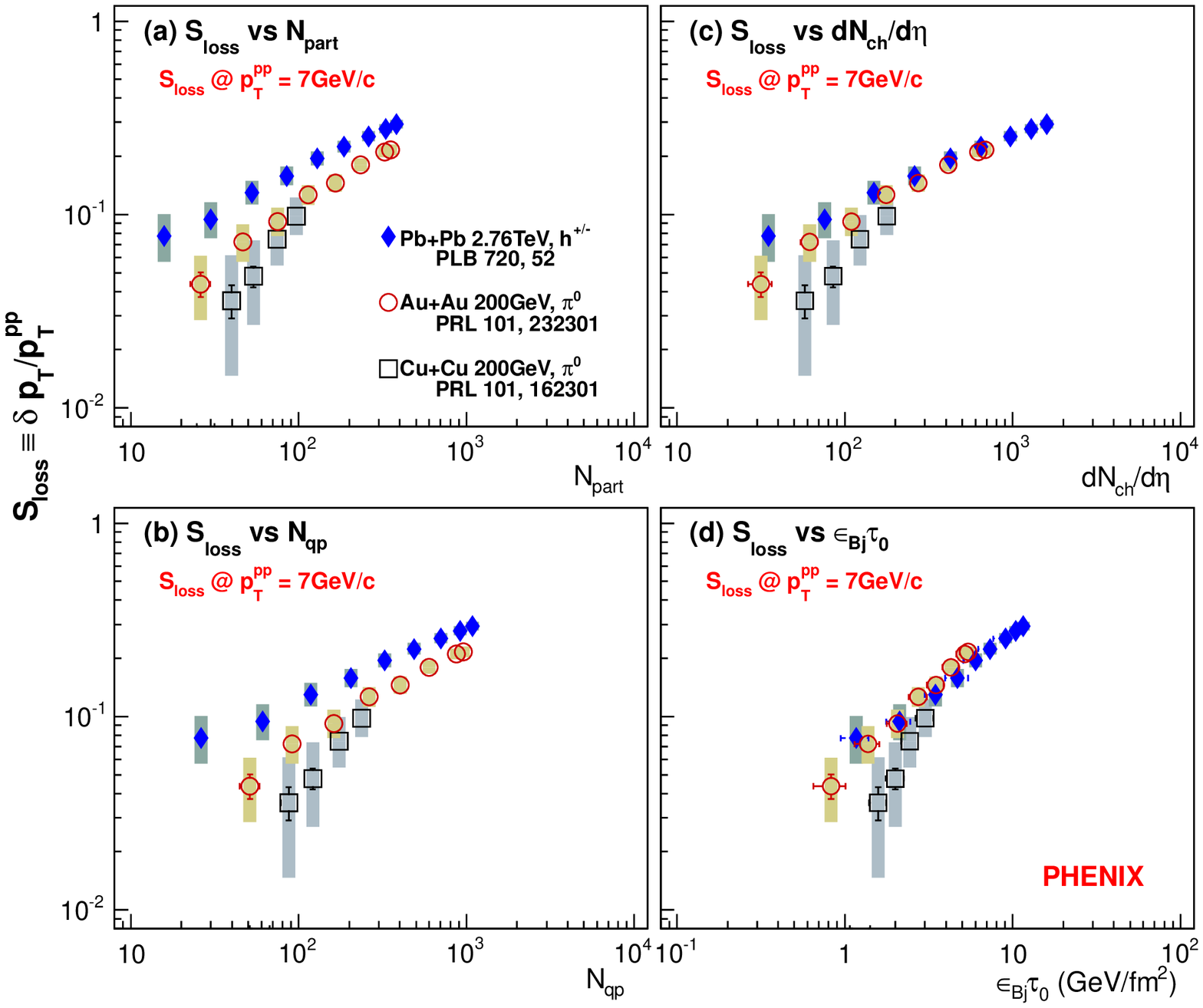}
\caption{(Color online) 
Scaling variables dependence of \sloss at \ptpp=~7~\gevc. (a) shows \sloss 
vs \Npart, (b) shows \sloss vs \Nqp, (c) shows \sloss vs \dNdeta, and (d) 
shows \sloss vs \ebjtau. \Nqp are all calculated by PHENIX. $\delta_{\rm 
sys}$(\taa$\oplus$~pp norm) is not shown in these plots.
  }
    \label{fig:SlossScaling_pT7GeV}
%\end{figure*}
%%%%%%%%%%%%%%%%%%%%%%%%%%%%%%%%%%%% Fig_11
%\begin{figure*}[!htb]
  \includegraphics[width=0.78\linewidth]{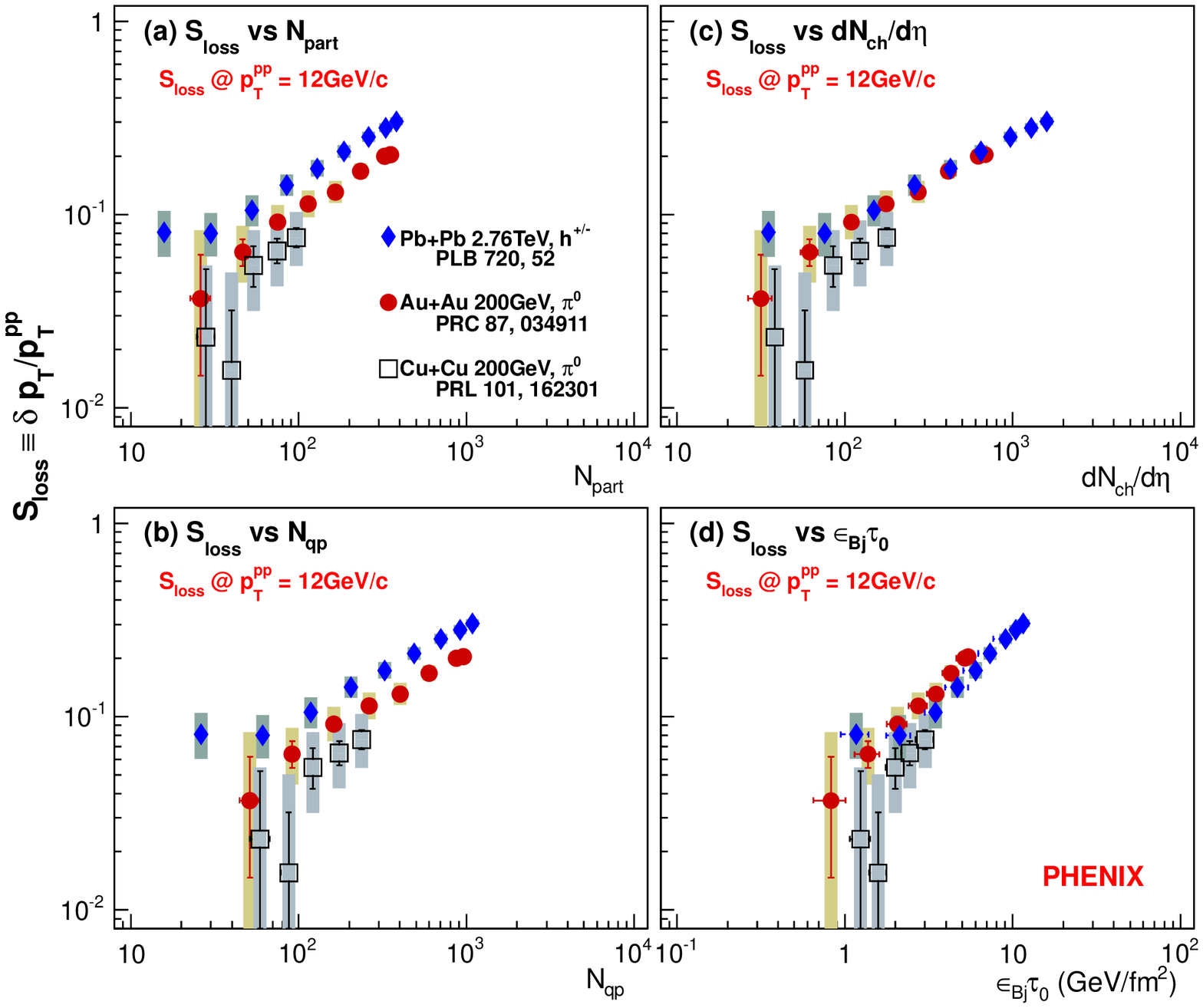}
\caption{(Color online) 
Scaling variables dependence of \sloss at \ptpp=~12~\gevc. (a) shows 
\sloss vs \Npart, (b) shows \sloss vs \Nqp, (c) shows \sloss vs \dNdeta, 
and (d) shows \sloss vs \ebjtau. \Nqp are all calculated by PHENIX. 
$\delta_{\rm sys}$(\taa$\oplus$~pp norm) is not shown in these plots.
  }
    \label{fig:SlossScaling_pT12GeV}
\end{figure*}

\begin{figure*}[!htb]
  \includegraphics[width=0.65\linewidth]{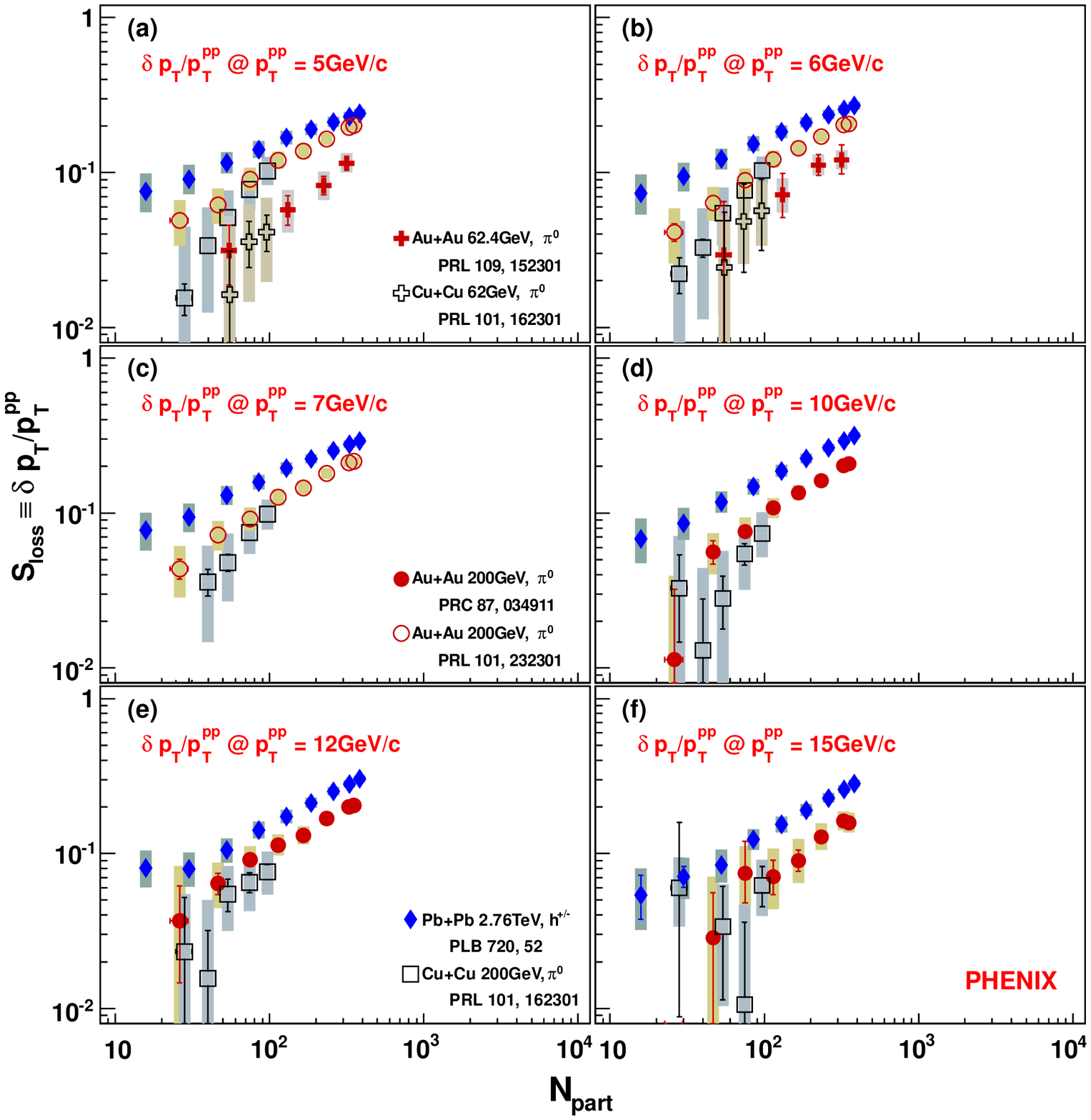}
  \caption{(Color online) \Npart dependence of the fractional momentum
  loss in bins of \ptpp for various systems and \sqsn.
$\delta_{\rm sys}$(\taa$\oplus$~pp norm) is not shown in these plots.
  }
    \label{fig:two-side-by-side-npart}
%\end{figure*}
%%%%%%%%%%%%%%%%%%%%%%%%%%%%%%%%%%%% Fig_13
%\begin{figure*}[!htb]
  \includegraphics[width=0.65\linewidth]{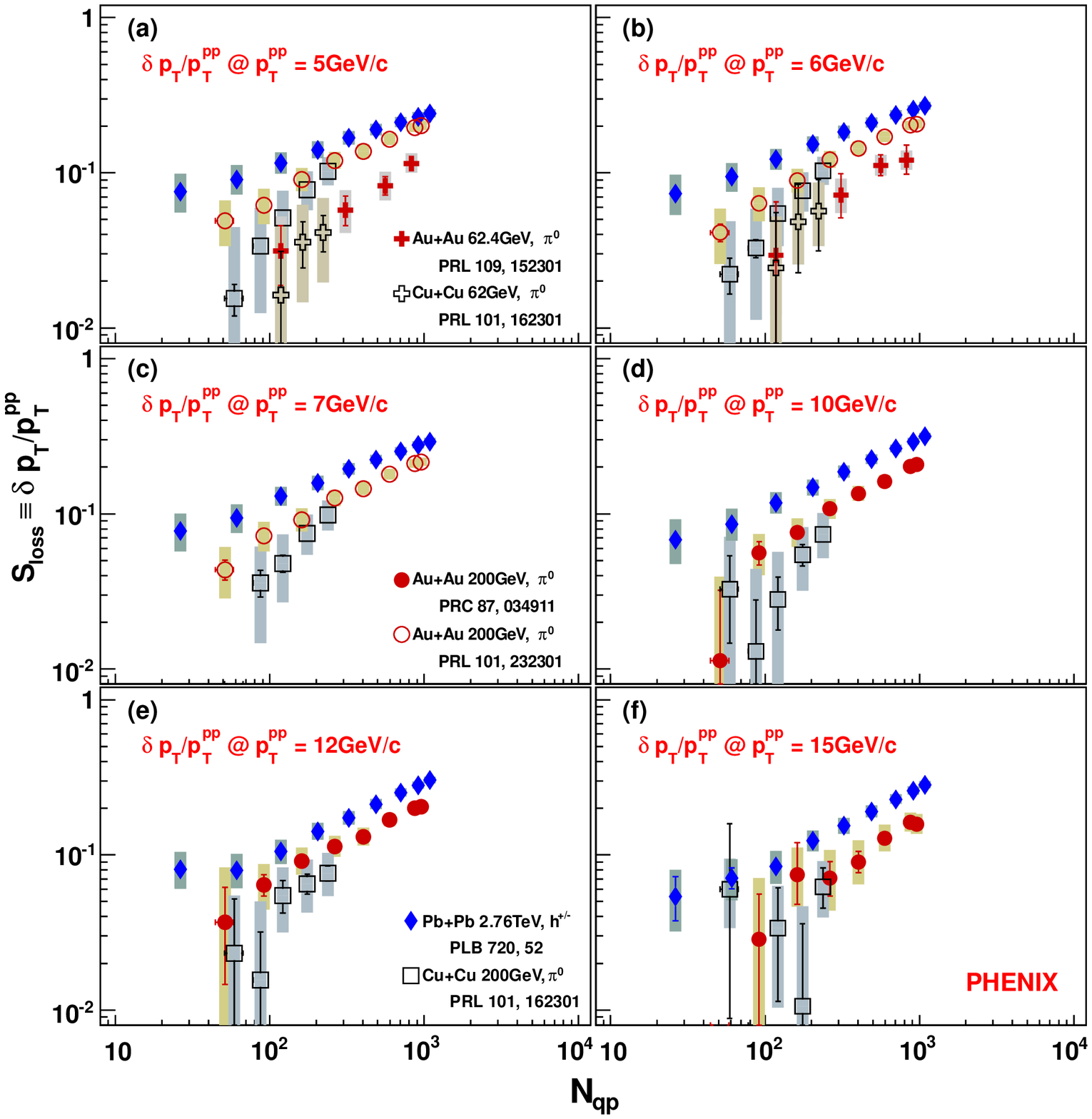}
          \caption{(Color online) \Nqp dependence of the fractional 
	  momentum loss in bins of \pt for various systems and \sqsn.
$\delta_{\rm sys}$(\taa$\oplus$~pp norm) is not shown in these plots.
\Nqp are all calculated by PHENIX.
  }
    \label{fig:two-side-by-side-nqp}
\end{figure*}

To understand how the fractional momentum loss changes with collision 
systems, we plot \sloss against the scaling variables defined in the 
section~\ref{AnaSection}.  Figures~\ref{fig:SlossScaling_pT7GeV} 
and~\ref{fig:SlossScaling_pT12GeV} show the \sloss as a function of 
\Npart, \Nqp, \dNdeta, and \ebjtau at \ptpp=~7 and 12~\gevc, respectively. 
Note that at these \ptpp values, only data from 200 GeV and 2.76 TeV are 
available.
When a value at the exact \ptpp was not available, we 
interpolated the fractional momentum loss from the closest two \pt points 
that we obtained in the previous section. The error bars represent Type A 
and the boxes are Type B uncertainties; Type C uncertainties are not shown 
here.  The scaling variable dependencies show clearer power-law behavior 
at \pt=~12~GeV/$c$ than at \pt=~7~GeV/$c$, implying that the Sloss is 
dominated by a single source, i.e., hard scattering. At fixed \sqsn, the 
\sloss values for the \cucu and \auau systems converge as \Npart grows.  
For the different \sqsn values, a clear separation of \sloss values is 
seen even at the highest \Npart, and the separation increases with 
increasing \pt (see Fig.~\ref{fig:two-side-by-side-npart}).

\begin{figure*}[!htb]
  \includegraphics[width=0.65\linewidth]{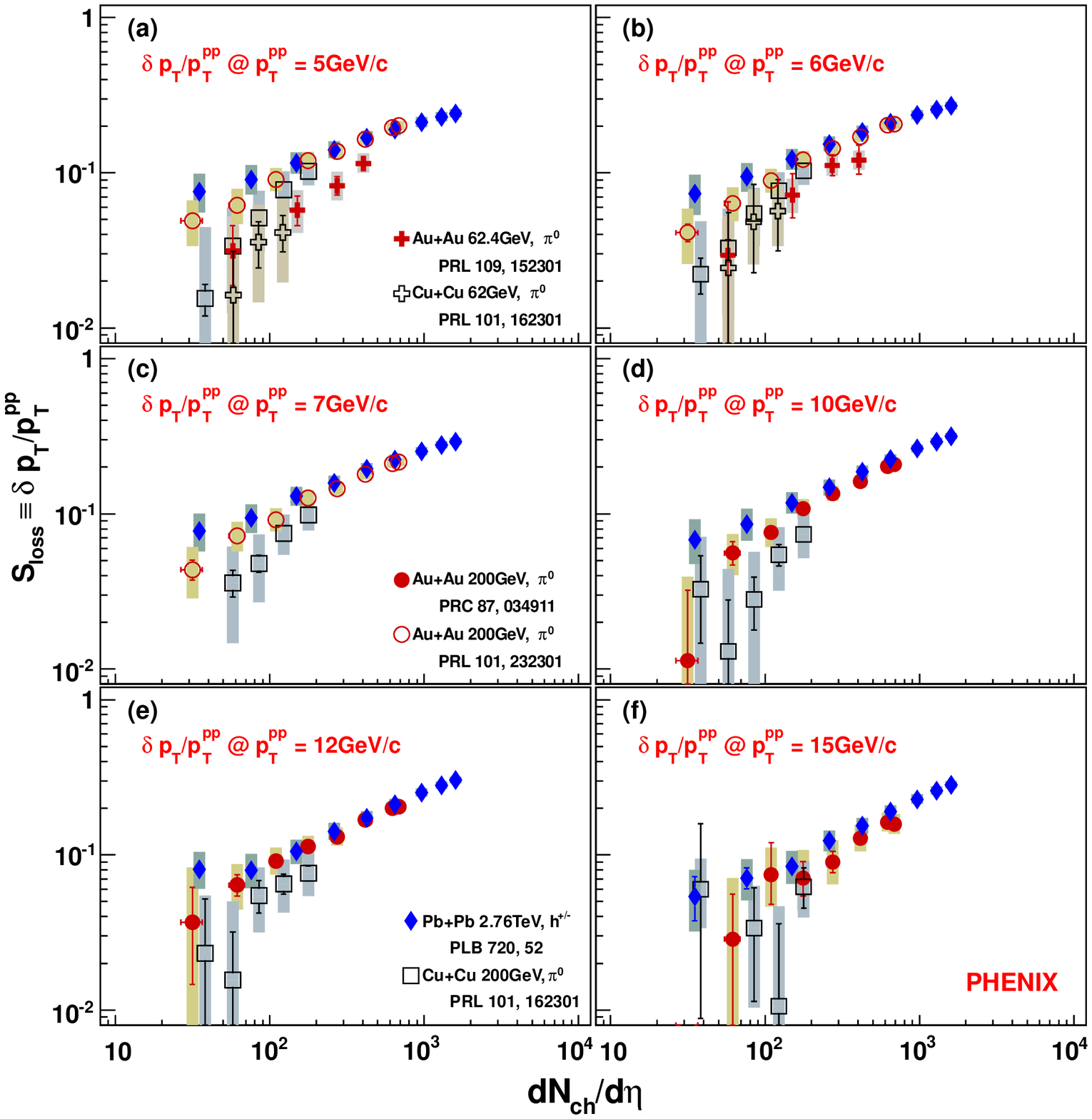}
          \caption{(Color online) \dNdeta dependence of the fractional momentum loss.
$\delta_{\rm sys}$(\taa$\oplus$~pp norm) is not shown in these plots.
  }
    \label{fig:two-side-by-side-dndeta}
%\end{figure*}
%%%%%%%%%%%%%%%%%%%%%%%%%%%%%%%%%%%% Fig_15
%\begin{figure*}[!htb]
  \includegraphics[width=0.65\linewidth]{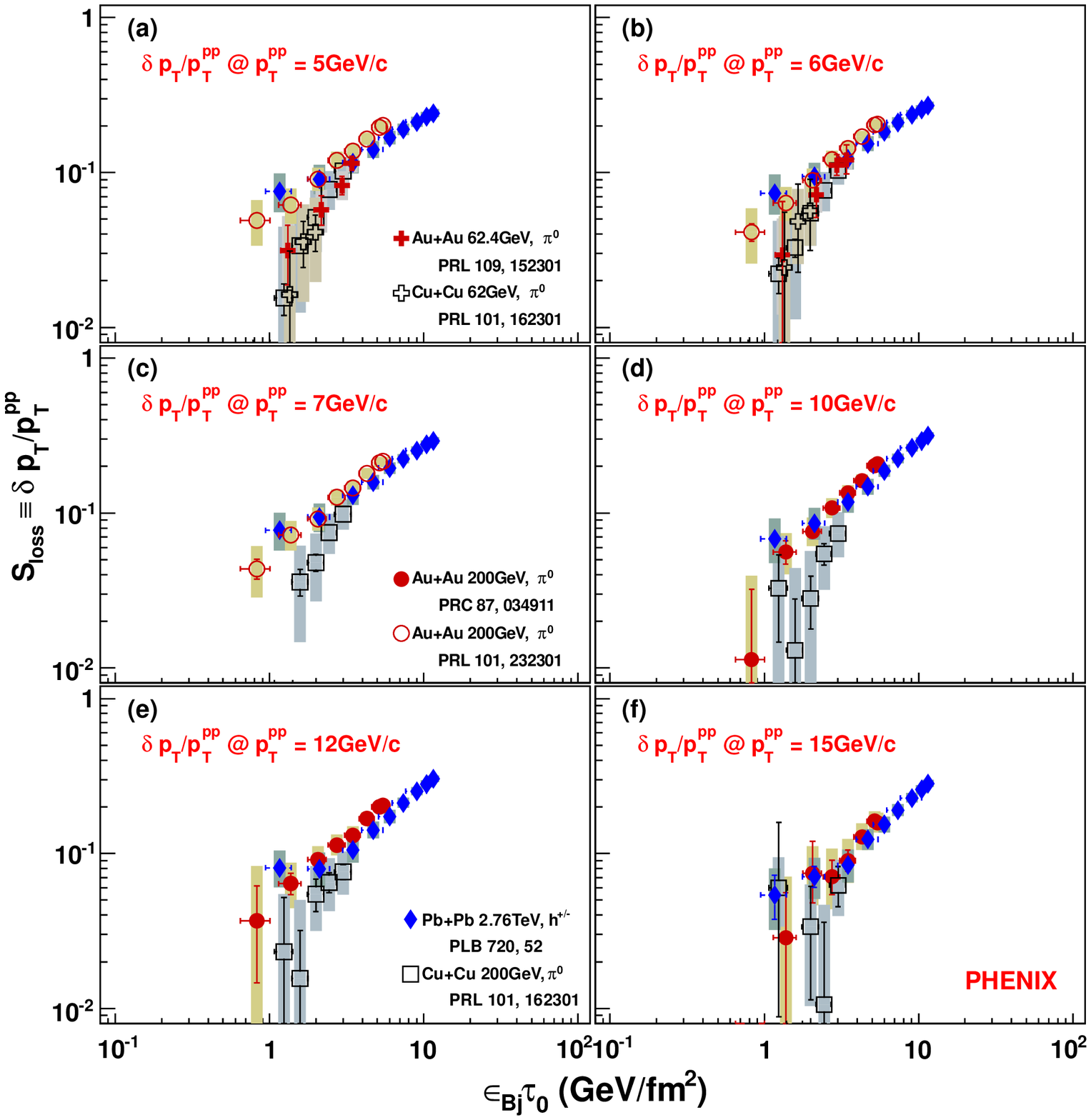}
          \caption{(Color online) \ebjtau dependence of the fractional momentum loss.
$\delta_{\rm sys}$(\taa$\oplus$~pp norm) is not shown in these plots.
  }
    \label{fig:two-side-by-side-ebj}
\end{figure*}

Figures~\ref{fig:two-side-by-side-npart}--\ref{fig:two-side-by-side-ebj}
show the same \sloss dependencies for additional \ptpp values of
5--15~\gevc.
For the lowest two \ptpp values, the results now also 
include Cu$+$Cu and Au$+$Au at \sqsn=~62.4~GeV.
Note that the PHENIX 
and ALICE data show parallel trends as a function of \Npart, especially at 
higher \Npart.  This fact, albeit the magnitudes are different, can be 
associated with the observation that ALICE and PHENIX data exhibit a 
similar \Npart dependence of the $\dNdeta/(0.5\Npart)$ 
shapes~\cite{Abelev:2012hxa}. When looking at \Nqp dependence, as expected 
from the discussion in the section explaining \Nqp, the points are shifted 
up by a factor of 2-3 along the x-axis. The overall trends are similar as 
for \Npart dependence, but the slopes are somewhat different. Comparing 
the data from different collision systems at the same \sqsn reveals no 
significant improvement of the alignment from \Npart to \Nqp scaling.
When we plot the \sloss against \dNdeta, the situation is different. 

At higher centralities (increasing \dNdeta) the LHC points line up very 
well with the 200~GeV RHIC \auau data, moreover, at higher \pt the two 
results are consistent for all but the most peripheral collisions. 
This clearly shows that \sloss scales with \dNdeta, which 
is energy density dependent and thus \sqsn dependent.  Finally, 
plots of \sloss as a function of 
\ebjtau~\ref{fig:two-side-by-side-ebj} show 
remarkable universal trends for the data from different systems from 
200~GeV to 2.76~TeV.  Among the scaling variables, \dNdeta and \ebjtau 
seems to serve best across the collision systems, especially between 
200~GeV Au$+$Au and 2.76~TeV collisions.  This investigation shows that 
the \sloss does not scale with simple geometry descriptions across the 
\sqsn, but do scale with the quantities related to the energy density of 
the system, hence the opacity of the system is energy-density dependent.

We have investigated \sloss against the four scaling variables at six 
\ptpp points including the two already shown in 
Figs.~\ref{fig:SlossScaling_pT7GeV} and~\ref{fig:SlossScaling_pT12GeV}. 
The scaling plots at all \ptpp 
are shown in Figs.~\ref{fig:two-side-by-side-npart} 
--~\ref{fig:two-side-by-side-ebj}.  For \pt of 5 and 6~\gevc, we used 
the 2004 data, because the 2007 data has a software threshold in \pt, 
as mentioned earlier.  At the same two lowest \pt, we also show the 
\sloss scaling for 62.4~GeV \cucu and \auau collisions. For higher 
\pt the 62.4~GeV points are not available owing to the lack of a \pp 
baseline. Deviations seen in the 62.4~GeV data may indicate that in 
the measured \pt range hard scattering is not completely dominant 
yet, in accordance with the observations of~\cite{Adare:2012uk}.

Lastly, to quantify the scaling trends, we fit \sloss for all 
four scaling variables and each collision system, except for 
\snn=~62.4~GeV system, with a power-law function:
\begin{equation}
\dptpt = \beta (SV/SV^{0})^{\alpha}
\label{eqfit1}
\end{equation}
where $SV$ is one of the four scaling variables we used above, and the 
$SV^{0}$ is the normalization factor introduced to cancel the dimension of 
the $SV$. We took the scaling variables for the most central LHC points as 
$SV^{0}$. Use of the power-law function is motivated by an energy loss 
model that predicts that $\Delta 
E/E\propto\Npart^{2/3}$~\cite{Vitev:2005he}. In the fitting process the 
statistical and systematic uncertainties were taken into account according 
to the prescription of~\cite{Adare:2008cg}. The errors on the scaling 
variable (horizontal errors in the plots) are not taken into account in 
the fitting, but they are small compared to the uncertainties of \sloss 
values.  

\begin{figure}[!htb]
  \includegraphics[width=1.0\linewidth]{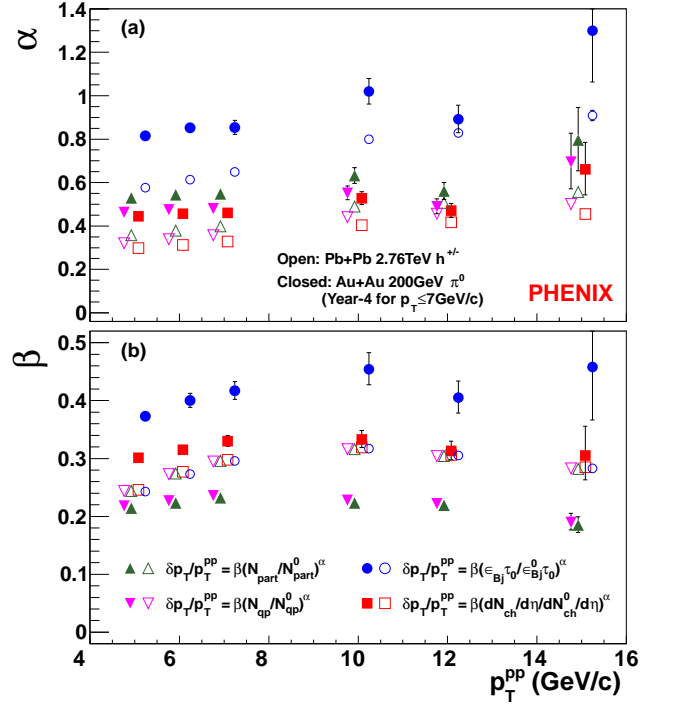}
\caption{(Color online) 
\ptpp dependence of fitting parameters for \sloss vs scaling variables. 
Open symbols correspond to Pb$+$Pb 2.76~TeV, and closed symbols correspond 
to Au$+$Au 200~GeV. (a) $\alpha$ and vs \ptpp (b) $\beta$ vs \ptpp. The 
Cu+Cu 200~GeV points are not shown, instead are tabulated in 
Table~\ref{FitParTablexmode}.
  }
    \label{fig:SlossFitParNpart}
\end{figure}

The fit parameters $\alpha$ and $\beta$ obtained by fitting \dptpt vs 
\Npart and \Nqp, plus \dNdeta and \ebjtau to Eq.~\ref{eqfit1} for Au$+$Au 
at \sqsn=~200~GeV and Pb$+$Pb at \sqsn=~2.76~TeV are shown in 
Fig.~\ref{fig:SlossFitParNpart}. All fit parameters, including for 
Cu$+$Cu, are tabulated in Table~\ref{FitParTablexmode}.

%%%%%%%%%%%%%%%%%%%%%%%%%%%%%%%%%%%% Fig_12

%%%%%%%%%%%%%%%%%%%%%%%%%%%%%%%%%%%% Fig_14

The fit parameters $\alpha$ and $\beta$ are anti-correlated.  At 
and above 10~\gevc, the $\chi^2/ndf$ values become smaller and 
the powers $\alpha$
converge for all scaling variables, although they do not become fully 
consistent within uncertainties. Among the scaling variables, \dNdeta is 
found to give relatively consistent $\alpha$ and $\beta$ between two 
systems. The \ebjtau, which is more related to the energy density of the 
system, also gives reasonably consistent numbers within
uncertainties. More interestingly, \ebjtau gives the $\alpha$ 
closest to 1.0 (linear scaling).  The similarities are striking as is 
the fact that \sloss obeys such a simple scaling with global observables 
over the entire \pt range where hard scattering is dominant. This implies 
that the empirical fractional momentum loss and the assumed underlying 
energy loss of partons scale with energy density of the medium, 
independent of the collision energies or systems, once \sqsn is 
sufficiently high. We cross-checked our current result with one published 
earlier for a slightly different quantity~\cite{Adare:2008qa}, and found 
consistent for \sqsn=~200~GeV Au$+$Au collisions.

\clearpage

		\section{Summary}
                \label{GrandSummary}

We have studied fractional momentum loss (\sloss$\equiv$\dptpt) over 
various systems and collision energies as a function of \pt and four 
scaling variables: \Npart, \Nqp, \dNdeta and \ebjtau. We found that the 
same universal function of \dNdeta or \ebjtau describes \sloss at RHIC 
(\sqsn=~200~GeV) and LHC (\sqsn=~2.76~TeV), while \Npart and \Nqp do not. 
This finding shows that the \sloss does not scale simply with system size 
across the \sqsn, but does scale with quantities related to the energy 
density of the system, implying that the opacity of the system is 
energy-density dependent.  We quantitatively evaluated the slope of the 
universal curves for \sqsn=~200 and 2.76~TeV and again found that \dNdeta 
and \ebjtau give relatively consistent $\alpha$ and $\beta$ between two 
systems, and especially, that the the $\alpha$ for \ebjtau is close to 1.0 
(linear scaling). It is striking that \sloss obeys such a simple scaling 
with global observables over the entire \pt range where hard scattering is 
dominant. This implies that the empirical fractional momentum loss and the 
assumed underlying energy loss of partons scale with energy density of the 
medium, independent of the collision energies or systems, once \sqsn is 
sufficiently high.

We propose that measurements of \sloss as well as the conventional \raa, 
in the future, would provide important additional information to 
investigate the global feature of the energy loss of partons.

%%%%%%%%%%%%%%%%%%%%%%%%%%%%%%%%%%%% Fig_16

%%%%%%%%%%%%%%%%%%%%%%%%%%%%%%%%%%%% Acknowledgments

\section*{ACKNOWLEDGMENTS}   % Run-14 long form for all journals

We thank the staff of the Collider-Accelerator and Physics
Departments at Brookhaven National Laboratory and the staff of
the other PHENIX participating institutions for their vital
contributions.  We acknowledge support from the
Office of Nuclear Physics in the
Office of Science of the Department of Energy,
the National Science Foundation,
Abilene Christian University Research Council,
Research Foundation of SUNY, and
Dean of the College of Arts and Sciences, Vanderbilt University
(U.S.A),
Ministry of Education, Culture, Sports, Science, and Technology
and the Japan Society for the Promotion of Science (Japan),
Conselho Nacional de Desenvolvimento Cient\'{\i}fico e
Tecnol{\'o}gico and Funda\c c{\~a}o de Amparo {\`a} Pesquisa do
Estado de S{\~a}o Paulo (Brazil),
Natural Science Foundation of China (P.~R.~China),
Croatian Science Foundation and
Ministry of Science, Education, and Sports (Croatia),
Ministry of Education, Youth and Sports (Czech Republic),
Centre National de la Recherche Scientifique, Commissariat
{\`a} l'{\'E}nergie Atomique, and Institut National de Physique
Nucl{\'e}aire et de Physique des Particules (France),
Bundesministerium f\"ur Bildung und Forschung, Deutscher
Akademischer Austausch Dienst, and Alexander von Humboldt Stiftung 
(Germany),
National Science Fund, OTKA, K\'aroly R\'obert University College,
and the Ch. Simonyi Fund (Hungary),
Department of Atomic Energy and Department of Science and Technology 
(India),
Israel Science Foundation (Israel),
Basic Science Research Program through NRF of the Ministry of Education 
(Korea),
Physics Department, Lahore University of Management Sciences (Pakistan),
Ministry of Education and Science, Russian Academy of Sciences,
Federal Agency of Atomic Energy (Russia),
VR and Wallenberg Foundation (Sweden),
the U.S. Civilian Research and Development Foundation for the
Independent States of the Former Soviet Union,
the Hungarian American Enterprise Scholarship Fund,
and the US-Israel Binational Science Foundation.

%%%%%%%%%%%%%%%%%%%%%%%%%%%%%%%%%%%% Appendix

\section*{APPENDIX}

Tables of the centrality dependence of $\delta p_T/p_T^{pp}$ and 
parameters for fitting four different power-law functions for Au$+$Au and 
Cu$+$Cu data from the PHENIX experiment at RHIC and Pb$+$Pb data from the 
ALICE experiment at the 
LHC~\protect\cite{Aamodt:2010jd,Abelev:2012hxa,Abelev:2014laa}.

%\clearpage

%\setcounter{table}{0} \renewcommand{\thetable}{A.\arabic{table}} 

%--------------------------------------------  Table_IV
\begin{table*}[!htb]
\caption{Centrality dependence of $\delta p_T/p_T^{pp}$ 
in Au$+$Au collisions at  $\sqrt{s_{_{NN}}}=200$~GeV 
from 2007 and 2004 data from the PHENIX experiment at RHIC.}
\begin{ruledtabular} \begin{tabular}{cccccccccc}
 &   \multicolumn{4}{c}{2007 data} && \multicolumn{4}{c}{2004 data} \\
Centrality & 
     $p_T^{pp}$ [GeV/$c$] & $\delta p_T/p_T^{pp}$ & Stat error & Syst error 
 & & $p_T^{pp}$ [GeV/$c$] & $\delta p_T/p_T^{pp}$ & Stat error & Syst error \\
\hline
0\%--5\% 
 & 7.0 & 0.216 & $_{-0.004}^{+0.004}$ & $_{-0.013}^{+0.015}$ &
 & 5.0 & 0.202 & $_{-0.003}^{+0.003}$ & $_{-0.013}^{+0.015}$ 
\\
 & 10.0 & 0.209 & $_{-0.005}^{+0.005}$ & $_{-0.014}^{+0.016}$ &
 & 6.0 & 0.206 & $_{-0.003}^{+0.004}$ & $_{-0.013}^{+0.015}$ 
\\
 & 12.0 & 0.204 & $_{-0.006}^{+0.007}$ & $_{-0.013}^{+0.016}$ &
 & 7.0 & 0.216 & $_{-0.002}^{+0.002}$ & $_{-0.013}^{+0.015}$ 
\\
 & 15.0 & 0.157 & $_{-0.010}^{+0.012}$ & $_{-0.021}^{+0.026}$ 
 & & & & 
\\ \\
0\%--10\%
 & 7.0 & 0.210 & $_{-0.001}^{+0.001}$ & $_{-0.013}^{+0.016}$ &
 & 5.0 & 0.196 & $_{-0.002}^{+0.002}$ & $_{-0.013}^{+0.015}$ 
\\
 & 10.0 & 0.202 & $_{-0.004}^{+0.004}$ & $_{-0.014}^{+0.016}$ &
 & 6.0 & 0.202 & $_{-0.002}^{+0.002}$ & $_{-0.013}^{+0.015}$ 
\\
 & 12.0 & 0.200 & $_{-0.005}^{+0.006}$ & $_{-0.013}^{+0.016}$ &
 & 7.0 & 0.211 & $_{-0.003}^{+0.003}$ & $_{-0.013}^{+0.015}$ 
\\ 
 & 15.0 & 0.162 & $_{-0.009}^{+0.010}$ & $_{-0.020}^{+0.026}$ &
 & & & & 
\\ \\
10\%--20\%
 & 7.0 & 0.172 & $_{-0.001}^{+0.001}$ & $_{-0.014}^{+0.016}$ &
 & 5.0 & 0.165 & $_{-0.002}^{+0.002}$ & $_{-0.014}^{+0.016}$ 
\\
 & 10.0 & 0.162 & $_{-0.005}^{+0.005}$ & $_{-0.014}^{+0.016}$ &
 & 6.0 & 0.171 & $_{-0.002}^{+0.002}$ & $_{-0.013}^{+0.015}$ 
\\
 & 12.0 & 0.168 & $_{-0.006}^{+0.007}$ & $_{-0.014}^{+0.017}$ &
 & 7.0 & 0.180 & $_{-0.003}^{+0.003}$ & $_{-0.013}^{+0.015}$ 
\\
 & 15.0 & 0.128 & $_{-0.011}^{+0.012}$ & $_{-0.022}^{+0.029}$ 
 & & & & 
\\ \\
20\%--30\%
 & 7.0 & 0.140 & $_{-0.002}^{+0.002}$ & $_{-0.015}^{+0.017}$ &
 & 5.0 & 0.137 & $_{-0.002}^{+0.002}$ & $_{-0.014}^{+0.016}$ 
\\
 & 10.0 & 0.135 & $_{-0.005}^{+0.006}$ & $_{-0.014}^{+0.016}$ &
 & 6.0 & 0.144 & $_{-0.002}^{+0.003}$ & $_{-0.014}^{+0.016}$ 
\\
 & 12.0 & 0.131 & $_{-0.006}^{+0.006}$ & $_{-0.016}^{+0.019}$ &
 & 7.0 & 0.145 & $_{-0.003}^{+0.003}$ & $_{-0.014}^{+0.016}$ 
\\
 & 15.0 & 0.090 & $_{-0.014}^{+0.016}$ & $_{-0.026}^{+0.034}$ &
 & & & &
\\ \\
30\%--40\%
 & 7.0 & 0.110 & $_{-0.002}^{+0.002}$ & $_{-0.015}^{+0.018}$ &
 & 5.0 & 0.120 & $_{-0.002}^{+0.002}$ & $_{-0.014}^{+0.016}$ 
\\
 & 10.0 & 0.108 & $_{-0.006}^{+0.006}$ & $_{-0.015}^{+0.017}$ &
 & 6.0 & 0.122 & $_{-0.003}^{+0.003}$ & $_{-0.014}^{+0.016}$ 
\\
 & 12.0 & 0.113 & $_{-0.007}^{+0.007}$ & $_{-0.016}^{+0.019}$ &
 & 7.0 & 0.126 & $_{-0.004}^{+0.004}$ & $_{-0.014}^{+0.016}$ 
\\
 & 15.0 & 0.071 & $_{-0.016}^{+0.020}$ & $_{-0.027}^{+0.037}$ &
 & & & &
\\ \\
40\%--50\%
 & 7.0 & 0.080 & $_{-0.002}^{+0.002}$ & $_{-0.016}^{+0.018}$ &
 & 5.0 & 0.091 & $_{-0.002}^{+0.002}$ & $_{-0.015}^{+0.017}$ 
\\
 & 10.0 & 0.076 & $_{-0.007}^{+0.008}$ & $_{-0.015}^{+0.017}$ &
 & 6.0 & 0.089 & $_{-0.003}^{+0.003}$ & $_{-0.015}^{+0.017}$ 
\\
 & 12.0 & 0.091 & $_{-0.007}^{+0.008}$ & $_{-0.017}^{+0.020}$ &
 & 7.0 & 0.092 & $_{-0.004}^{+0.004}$ & $_{-0.015}^{+0.017}$ 
\\
 & 15.0 & 0.075 & $_{-0.027}^{+0.045}$ & $_{-0.028}^{+0.037}$ &
 & & & &
\\ \\
50\%--60\%
 & 7.0 & 0.055 & $_{-0.003}^{+0.003}$ & $_{-0.016}^{+0.019}$ &
 & 5.0 & 0.062 & $_{-0.003}^{+0.003}$ & $_{-0.015}^{+0.017}$ 
\\
 & 10.0 & 0.056 & $_{-0.009}^{+0.010}$ & $_{-0.016}^{+0.018}$ &
 & 6.0 & 0.064 & $_{-0.004}^{+0.004}$ & $_{-0.015}^{+0.017}$ 
\\
 & 12.0 & 0.064 & $_{-0.010}^{+0.011}$ & $_{-0.019}^{+0.023}$ &
 & 7.0 & 0.072 & $_{-0.005}^{+0.005}$ & $_{-0.015}^{+0.017}$ 
\\
 & 15.0 & 0.029 & $_{-0.022}^{+0.027}$ & $_{-0.031}^{+0.042}$ &
 & & & &
\\ \\
60\%--70\%
 & 7.0 & 0.028 & $_{-0.004}^{+0.004}$ & $_{-0.017}^{+0.019}$ &
 & 5.0 & 0.049 & $_{-0.003}^{+0.003}$ & $_{-0.015}^{+0.017}$ 
\\
 & 10.0 & 0.011 & $_{-0.019}^{+0.021}$ & $_{-0.024}^{+0.028}$ &
 & 6.0 & 0.041 & $_{-0.005}^{+0.006}$ & $_{-0.015}^{+0.018}$ 
\\
 & 12.0 & 0.037 & $_{-0.022}^{+0.025}$ & $_{-0.037}^{+0.046}$ &
 & 7.0 & 0.044 & $_{-0.006}^{+0.007}$ & $_{-0.015}^{+0.018}$ 
\\
 & 15.0 & -0.098 & $_{-0.063}^{+0.046}$ & $_{-0.077}^{+0.053}$ &
 & & & &
\end{tabular} \end{ruledtabular}
\end{table*}

%--------------------------------------------  Table_V
\begin{table*}[!htb]
\caption{Centrality dependence of $\delta p_T/p_T^{pp}$ 
in Au$+$Au collisions at $\sqrt{s_{_{NN}}}=62.4$~GeV and 
Cu$+$Cu collisions at at $\sqrt{s_{_{NN}}}=200$ and 62.4~GeV
from the PHENIX experiment at RHIC.}
\begin{minipage}{0.48\linewidth}
\begin{ruledtabular} \begin{tabular}{cccccc}
 System & Centrality          & $p_T^{pp}$ & $\delta p_T/p_T^{pp}$  
& stat & syst \\
 $\sqrt{s_{_{NN}}}$ & &  [GeV/$c$] & 
& uncert. & uncert.\\
\hline
Au$+$Au & 0\%--10\%
 & 5.0 & 0.115 & $_{-0.009}^{+0.010}$ & $_{-0.015}^{+0.018}$ \\
62.4~GeV 
& & 6.0 & 0.120 & $_{-0.023}^{+0.030}$ & $_{-0.016}^{+0.019}$ \\
\\
 & 10\%--20\%
 & 5.0 & 0.083 & $_{-0.010}^{+0.012}$ & $_{-0.016}^{+0.019}$ \\
& & 6.0 & 0.112 & $_{-0.016}^{+0.019}$ & $_{-0.016}^{+0.019}$ \\
\\
 & 20\%--40\%
 & 5.0 & 0.057 & $_{-0.012}^{+0.013}$ & $_{-0.016}^{+0.020}$ \\
& & 6.0 & 0.072 & $_{-0.021}^{+0.027}$ & $_{-0.017}^{+0.020}$ \\
\\
Cu$+$Cu & 0\%--10\%
 & 5.0 & 0.102 & $_{-0.001}^{+0.001}$ & $_{-0.020}^{+0.024}$ \\
200~GeV
& & 6.0 & 0.103 & $_{-0.002}^{+0.002}$ & $_{-0.020}^{+0.024}$ \\
& & 7.0 & 0.098 & $_{-0.004}^{+0.004}$ & $_{-0.020}^{+0.024}$ \\
& & 10.0 & 0.074 & $_{-0.007}^{+0.008}$ & $_{-0.022}^{+0.027}$ \\
& & 12.0 & 0.076 & $_{-0.008}^{+0.009}$ & $_{-0.022}^{+0.027}$ \\
& & 15.0 & 0.062 & $_{-0.017}^{+0.020}$ & $_{-0.023}^{+0.029}$ \\
\\
 & 10\%--20\%
 & 5.0 & 0.078 & $_{-0.002}^{+0.002}$ & $_{-0.020}^{+0.024}$ \\
& & 6.0 & 0.077 & $_{-0.003}^{+0.003}$ & $_{-0.020}^{+0.024}$ \\
& & 7.0 & 0.075 & $_{-0.004}^{+0.005}$ & $_{-0.020}^{+0.025}$ \\
& & 10.0 & 0.054 & $_{-0.008}^{+0.009}$ & $_{-0.022}^{+0.028}$ \\
& & 12.0 & 0.065 & $_{-0.009}^{+0.010}$ & $_{-0.022}^{+0.028}$ \\
& & 15.0 & 0.011 & $_{-0.021}^{+0.025}$ & $_{-0.027}^{+0.036}$ \\
\\
 & 20\%--30\%
 & 5.0 & 0.051 & $_{-0.002}^{+0.002}$ & $_{-0.021}^{+0.025}$ \\
& & 6.0 & 0.054 & $_{-0.004}^{+0.004}$ & $_{-0.021}^{+0.025}$ \\
& & 7.0 & 0.048 & $_{-0.006}^{+0.006}$ & $_{-0.021}^{+0.026}$ \\
& & 10.0 & 0.028 & $_{-0.010}^{+0.011}$ & $_{-0.023}^{+0.029}$ \\
& & 12.0 & 0.055 & $_{-0.012}^{+0.014}$ & $_{-0.023}^{+0.029}$ \\
& & 15.0 & 0.034 & $_{-0.022}^{+0.028}$ & $_{-0.023}^{+0.029}$ \\
\end{tabular} \end{ruledtabular}
\end{minipage}
\begin{minipage}{0.48\linewidth}
\begin{ruledtabular} \begin{tabular}{cccccc}
 System & Centrality          & $p_T^{pp}$ & $\delta p_T/p_T^{pp}$  
& stat & syst \\
 $\sqrt{s_{_{NN}}}$ & &  [GeV/$c$] & 
& uncert. & uncert.\\
\hline
Cu$+$Cu 
& 30\%--40\%
 & 5.0 & 0.034 & $_{-0.002}^{+0.002}$ & $_{-0.021}^{+0.026}$ \\
200~GeV
& & 6.0 & 0.033 & $_{-0.004}^{+0.004}$ & $_{-0.021}^{+0.026}$ \\
(continued)
& & 7.0 & 0.036 & $_{-0.007}^{+0.007}$ & $_{-0.021}^{+0.026}$ \\
& & 10.0 & 0.013 & $_{-0.013}^{+0.015}$ & $_{-0.025}^{+0.031}$ \\
& & 12.0 & 0.016 & $_{-0.015}^{+0.016}$ & $_{-0.028}^{+0.035}$ \\
& & 15.0 & -0.001 & $_{-0.035}^{+0.028}$ & $_{-0.035}^{+0.028}$ \\
\\
& 40\%--50\%
 & 5.0 & 0.015 & $_{-0.004}^{+0.004}$ & $_{-0.024}^{+0.029}$ \\
& & 6.0 & 0.022 & $_{-0.006}^{+0.006}$ & $_{-0.022}^{+0.027}$ \\
& & 7.0 & -0.002 & $_{-0.016}^{+0.015}$ & $_{-0.042}^{+0.034}$ \\
& & 10.0 & 0.033 & $_{-0.018}^{+0.021}$ & $_{-0.031}^{+0.038}$ \\
& & 12.0 & 0.023 & $_{-0.023}^{+0.029}$ & $_{-0.025}^{+0.031}$ \\
& & 15.0 & 0.060 & $_{-0.051}^{+0.099}$ & $_{-0.026}^{+0.034}$ \\
\\ \\
Cu$+$Cu & 0\%--10\%
 & 5.0 & 0.041 & $_{-0.010}^{+0.012}$ & $_{-0.022}^{+0.028}$ \\
62.4~GeV 
& & 6.0 & 0.057 & $_{-0.025}^{+0.034}$ & $_{-0.023}^{+0.030}$ \\
\\
 & 10\%--20\%
 & 5.0 & 0.036 & $_{-0.011}^{+0.013}$ & $_{-0.021}^{+0.027}$ \\
& & 6.0 & 0.048 & $_{-0.026}^{+0.035}$ & $_{-0.023}^{+0.030}$ \\
\\
 & 20\%--30\%
 & 5.0 & 0.016 & $_{-0.013}^{+0.015}$ & $_{-0.022}^{+0.029}$ \\
& & 6.0 & 0.024 & $_{-0.024}^{+0.031}$ & $_{-0.022}^{+0.028}$ \\
\\
 & 30\%--40\%
 & 5.0 & 0.005 & $_{-0.024}^{+0.028}$ & $_{-0.044}^{+0.056}$ \\
& & 6.0 & -0.010 & $_{-0.163}^{+0.127}$ & $_{-0.180}^{+0.137}$ \\
\\
 & 40\%--50\%
 & 5.0 & -0.019 & $_{-0.021}^{+0.018}$ & $_{-0.033}^{+0.026}$ \\
& & 6.0 & -0.034 & $_{-0.050}^{+0.035}$ & $_{-0.024}^{+0.019}$ \\
\end{tabular} \end{ruledtabular}
\end{minipage}
\end{table*}

%--------------------------------------------  Table_VI
\begin{table*}[!htb]
\caption{Centrality dependence of $\delta p_T/p_T^{pp}$ 
Pb$+$Pb collisions at $\sqrt{s_{_{NN}}}=2.76$~TeV 
from the spectra measured by the ALICE experiment at the 
LHC~\protect\cite{Aamodt:2010jd,Abelev:2012hxa,Abelev:2014laa}.}
\begin{minipage}{0.48\linewidth}
\begin{ruledtabular} \begin{tabular}{ccccc}
Centrality & $p_T^{pp}$ [GeV/$c$] & $\delta p_T/p_T^{pp}$ & Stat error & Syst error \\
\hline
0\%--5\%
 & 5.0 & 0.241 & $_{-0.001}^{+0.001}$ & $_{-0.015}^{+0.017}$ \\
 & 6.0 & 0.270 & $_{-0.001}^{+0.001}$ & $_{-0.014}^{+0.016}$ \\
 & 7.0 & 0.293 & $_{-0.001}^{+0.001}$ & $_{-0.014}^{+0.015}$ \\
 & 10.0 & 0.316 & $_{-0.001}^{+0.001}$ & $_{-0.013}^{+0.015}$ \\
 & 12.0 & 0.303 & $_{-0.001}^{+0.001}$ & $_{-0.013}^{+0.015}$ \\
 & 15.0 & 0.282 & $_{-0.002}^{+0.002}$ & $_{-0.014}^{+0.016}$ \\
\\
5\%--10\%
 & 5.0 & 0.229 & $_{-0.001}^{+0.001}$ & $_{-0.015}^{+0.017}$ \\
 & 6.0 & 0.255 & $_{-0.001}^{+0.001}$ & $_{-0.014}^{+0.016}$ \\
 & 7.0 & 0.277 & $_{-0.001}^{+0.001}$ & $_{-0.014}^{+0.016}$ \\
 & 10.0 & 0.293 & $_{-0.001}^{+0.001}$ & $_{-0.014}^{+0.015}$ \\
 & 12.0 & 0.281 & $_{-0.002}^{+0.002}$ & $_{-0.014}^{+0.016}$ \\
 & 15.0 & 0.259 & $_{-0.002}^{+0.003}$ & $_{-0.015}^{+0.017}$ \\
\\
10\%--20\%
 & 5.0 & 0.211 & $_{-0.001}^{+0.001}$ & $_{-0.015}^{+0.017}$ \\
 & 6.0 & 0.236 & $_{-0.001}^{+0.001}$ & $_{-0.015}^{+0.017}$ \\
 & 7.0 & 0.253 & $_{-0.001}^{+0.001}$ & $_{-0.014}^{+0.016}$ \\
 & 10.0 & 0.263 & $_{-0.001}^{+0.001}$ & $_{-0.014}^{+0.016}$ \\
 & 12.0 & 0.252 & $_{-0.001}^{+0.002}$ & $_{-0.014}^{+0.016}$ \\
 & 15.0 & 0.228 & $_{-0.002}^{+0.002}$ & $_{-0.015}^{+0.018}$ \\
\\
20\%--30\%
 & 5.0 & 0.190 & $_{-0.001}^{+0.001}$ & $_{-0.015}^{+0.018}$ \\
 & 6.0 & 0.210 & $_{-0.001}^{+0.001}$ & $_{-0.015}^{+0.017}$ \\
 & 7.0 & 0.224 & $_{-0.001}^{+0.001}$ & $_{-0.015}^{+0.017}$ \\
 & 10.0 & 0.224 & $_{-0.001}^{+0.001}$ & $_{-0.015}^{+0.017}$ \\
 & 12.0 & 0.212 & $_{-0.002}^{+0.002}$ & $_{-0.015}^{+0.017}$ \\
 & 15.0 & 0.190 & $_{-0.003}^{+0.003}$ & $_{-0.016}^{+0.019}$ \\
\\
30\%--40\%
 & 5.0 & 0.168 & $_{-0.001}^{+0.001}$ & $_{-0.016}^{+0.018}$ \\
 & 6.0 & 0.183 & $_{-0.001}^{+0.001}$ & $_{-0.016}^{+0.018}$ \\
 & 7.0 & 0.195 & $_{-0.001}^{+0.001}$ & $_{-0.015}^{+0.018}$ \\
\end{tabular} \end{ruledtabular}
\end{minipage}
\begin{minipage}{0.48\linewidth}
\begin{ruledtabular} \begin{tabular}{ccccc}
Centrality & $p_T^{pp}$ [GeV/$c$] & $\delta p_T/p_T^{pp}$ & Stat error & Syst error \\
\hline
30\%--40\%
 & 10.0 & 0.187 & $_{-0.002}^{+0.002}$ & $_{-0.016}^{+0.018}$ \\
(continued)  
 & 12.0 & 0.173 & $_{-0.002}^{+0.002}$ & $_{-0.016}^{+0.018}$ \\
 & 15.0 & 0.154 & $_{-0.004}^{+0.004}$ & $_{-0.017}^{+0.020}$ \\
\\
40\%--50\%
 & 5.0 & 0.141 & $_{-0.001}^{+0.001}$ & $_{-0.017}^{+0.019}$ \\
 & 6.0 & 0.153 & $_{-0.001}^{+0.001}$ & $_{-0.016}^{+0.019}$ \\
 & 7.0 & 0.158 & $_{-0.001}^{+0.001}$ & $_{-0.016}^{+0.019}$ \\
 & 10.0 & 0.148 & $_{-0.002}^{+0.002}$ & $_{-0.017}^{+0.019}$ \\
 & 12.0 & 0.142 & $_{-0.003}^{+0.003}$ & $_{-0.017}^{+0.019}$ \\
 & 15.0 & 0.123 & $_{-0.005}^{+0.005}$ & $_{-0.018}^{+0.021}$ \\
\\
50\%--60\%
 & 5.0 & 0.116 & $_{-0.001}^{+0.001}$ & $_{-0.017}^{+0.020}$ \\
 & 6.0 & 0.122 & $_{-0.001}^{+0.001}$ & $_{-0.017}^{+0.020}$ \\
 & 7.0 & 0.130 & $_{-0.002}^{+0.002}$ & $_{-0.017}^{+0.020}$ \\
 & 10.0 & 0.118 & $_{-0.003}^{+0.003}$ & $_{-0.018}^{+0.020}$ \\
 & 12.0 & 0.105 & $_{-0.004}^{+0.004}$ & $_{-0.018}^{+0.021}$ \\
 & 15.0 & 0.084 & $_{-0.007}^{+0.007}$ & $_{-0.019}^{+0.022}$ \\
\\
60\%--70\%
 & 5.0 & 0.091 & $_{-0.002}^{+0.002}$ & $_{-0.019}^{+0.021}$ \\
 & 6.0 & 0.094 & $_{-0.002}^{+0.002}$ & $_{-0.019}^{+0.021}$ \\
 & 7.0 & 0.094 & $_{-0.002}^{+0.003}$ & $_{-0.019}^{+0.021}$ \\
 & 10.0 & 0.086 & $_{-0.004}^{+0.004}$ & $_{-0.019}^{+0.022}$ \\
 & 12.0 & 0.080 & $_{-0.006}^{+0.006}$ & $_{-0.019}^{+0.022}$ \\
 & 15.0 & 0.071 & $_{-0.010}^{+0.011}$ & $_{-0.020}^{+0.023}$ \\
\\
70\%--80\%
 & 5.0 & 0.075 & $_{-0.003}^{+0.003}$ & $_{-0.020}^{+0.023}$ \\
 & 6.0 & 0.074 & $_{-0.003}^{+0.003}$ & $_{-0.020}^{+0.024}$ \\
 & 7.0 & 0.077 & $_{-0.004}^{+0.004}$ & $_{-0.020}^{+0.023}$ \\
 & 10.0 & 0.068 & $_{-0.006}^{+0.006}$ & $_{-0.021}^{+0.024}$ \\
 & 12.0 & 0.081 & $_{-0.009}^{+0.010}$ & $_{-0.020}^{+0.024}$ \\
 & 15.0 & 0.054 & $_{-0.017}^{+0.019}$ & $_{-0.022}^{+0.026}$ \\
\end{tabular} \end{ruledtabular}
\end{minipage}
\end{table*}

%--------------------------------------------  Table_VII
\begin{table*}[!htb]
\caption{Parameters from fitting the indicated power-law functions 
for \dptpt to the data as a function of \ptpp for Au$+$Au 
collisions from 2004 and 2007 data and for Cu$+$Cu
collisions from 2005 data at \sqsn=~200~GeV.
}
\label{FitParTablexmode}
\begin{ruledtabular} \begin{tabular}{ccccccccccc}
System & \sqsn & year & hadron & \dptpt$=$ 
& \ptpp & $\alpha$ & $\beta$ & $\chi^{2}/ndf$ \\
\hline
%%%% A A A
Au$+$Au & 200~GeV & 2004 & $\pi^0$ & $\beta (\Npart/N^{0}_{\rm part})^{\alpha}$ 
 &  5~GeV/$c$ & $0.529_{-0.011}^{+0.011}$ & $2.14_{-0.03}^{+0.04}\times10^{-1}$ & 25.45/5 \\
 & & & &
 &  6~GeV/$c$ & $0.543_{-0.015}^{+0.015}$ & $2.23_{-0.04}^{+0.04}\times10^{-1}$ & 15.56/5 \\
 & & & &
 &  7~GeV/$c$ & $0.548_{-0.020}^{+0.020}$ & $2.32_{-0.04}^{+0.05}\times10^{-1}$ & 7.11/5 \\
       &         &      &         & $\beta (\Nqp/N^{0}_{\rm qp})^{\alpha}$ 
 &  5~GeV/$c$ & $0.463_{-0.010}^{+0.010}$ & $2.18_{-0.04}^{+0.04}\times10^{-1}$ & 23.35/5 \\
 & & & &
 &  6~GeV/$c$ & $0.475_{-0.013}^{+0.013}$ & $2.27_{-0.04}^{+0.04}\times10^{-1}$ & 15.23/5 \\
 & & & &
 &  7~GeV/$c$ & $0.480_{-0.017}^{+0.017}$ & $2.36_{-0.05}^{+0.05}\times10^{-1}$ & 7.50/5 \\
       &         &      &         & $\beta (dN_{ch}/d\eta/dN^{0}_{ch}/d\eta)^{\alpha}$ 
 &  5~GeV/$c$ & $0.445_{-0.009}^{+0.009}$ & $3.01_{-0.06}^{+0.06}\times10^{-1}$ & 27.78/5 \\
 & & & &
 &  6~GeV/$c$ & $0.456_{-0.013}^{+0.013}$ & $3.15_{-0.07}^{+0.08}\times10^{-1}$ & 18.56/5 \\
 & & & &
 &  7~GeV/$c$ & $0.460_{-0.016}^{+0.017}$ & $3.30_{-0.09}^{+0.10}\times10^{-1}$ & 8.50/5 \\
       &         &      &         & $\beta (\epsilon\tau_{0}/\epsilon^{0}\tau_{0})^{\alpha}$ 
 &  5~GeV/$c$ & $0.815_{-0.018}^{+0.018}$ & $3.73_{-0.09}^{+0.09}\times10^{-1}$ & 14.67/5 \\
 & & & &
 &  6~GeV/$c$ & $0.852_{-0.025}^{+0.025}$ & $4.00_{-0.12}^{+0.12}\times10^{-1}$ & 3.79/5 \\
 & & & &
 &  7~GeV/$c$ & $0.854_{-0.032}^{+0.032}$ & $4.17_{-0.15}^{+0.16}\times10^{-1}$ & 4.23/5 \\

Au$+$Au & 200~GeV & 2007 & $\pi^0$ & $\beta (\Npart/N^{0}_{\rm part})^{\alpha}$ 
 & 10~GeV/$c$ & $0.632_{-0.035}^{+0.036}$ & $2.23_{-0.06}^{+0.06}\times10^{-1}$ & 3.31/5 \\
 & & & &
 & 12~GeV/$c$ & $0.561_{-0.038}^{+0.040}$ & $2.19_{-0.07}^{+0.07}\times10^{-1}$ & 1.75/5 \\
 & & & &
 & 15~GeV/$c$ & $0.795_{-0.141}^{+0.151}$ & $1.85_{-0.13}^{+0.14}\times10^{-1}$ & 4.68/5 \\
       &         &      &         & $\beta (\Nqp/N^{0}_{\rm qp})^{\alpha}$ 
 & 10~GeV/$c$ & $0.552_{-0.031}^{+0.032}$ & $2.28_{-0.06}^{+0.06}\times10^{-1}$ & 3.32/5 \\
 & & & &
 & 12~GeV/$c$ & $0.490_{-0.034}^{+0.035}$ & $2.22_{-0.07}^{+0.07}\times10^{-1}$ & 1.78/5 \\
 & & & &
 & 15~GeV/$c$ & $0.695_{-0.124}^{+0.132}$ & $1.90_{-0.14}^{+0.15}\times10^{-1}$ & 4.74/5 \\
       &         &      &         & $\beta (dN_{ch}/d\eta/dN^{0}_{ch}/d\eta)^{\alpha}$ 
 & 10~GeV/$c$ & $0.528_{-0.029}^{+0.030}$ & $3.33_{-0.14}^{+0.15}\times10^{-1}$ & 3.72/5 \\
 & & & &
 & 12~GeV/$c$ & $0.471_{-0.032}^{+0.033}$ & $3.13_{-0.15}^{+0.17}\times10^{-1}$ & 1.59/5 \\
 & & & &
 & 15~GeV/$c$ & $0.661_{-0.117}^{+0.124}$ & $3.05_{-0.42}^{+0.51}\times10^{-1}$ & 4.69/5 \\
       &         &      &         & $\beta (\epsilon\tau_{0}/\epsilon^{0}\tau_{0})^{\alpha}$ 
 & 10~GeV/$c$ & $1.020_{-0.058}^{+0.060}$ & $4.54_{-0.26}^{+0.29}\times10^{-1}$ & 2.05/5 \\
 & & & &
 & 12~GeV/$c$ & $0.892_{-0.063}^{+0.064}$ & $4.05_{-0.27}^{+0.29}\times10^{-1}$ & 2.43/5 \\
 & & & &
 & 15~GeV/$c$ & $1.300_{-0.237}^{+0.255}$ & $4.58_{-0.91}^{+1.23}\times10^{-1}$ & 4.36/5 \\

Cu$+$Cu & 200~GeV & 2005 & $\pi^0$ & $\beta (\Npart/N^{0}_{\rm part})^{\alpha}$ 
 &  5~GeV/$c$ & $1.210_{-0.045}^{+0.046}$ & $5.45_{-0.37}^{+0.41}\times10^{-1}$ & 8.28/3 \\
 & & & &
 &  6~GeV/$c$ & $1.180_{-0.079}^{+0.082}$ & $5.21_{-0.60}^{+0.70}\times10^{-1}$ & 1.48/3 \\
 & & & &
 &  7~GeV/$c$ & $1.200_{-0.141}^{+0.148}$ & $5.17_{-1.01}^{+1.31}\times10^{-1}$ & 2.92/3 \\
       &         &      &         & $\beta (\Nqp/N^{0}_{\rm qp})^{\alpha}$ 
 &  5~GeV/$c$ & $1.060_{-0.039}^{+0.040}$ & $5.21_{-0.35}^{+0.38}\times10^{-1}$ & 9.71/3 \\
 & & & &
 &  6~GeV/$c$ & $1.030_{-0.069}^{+0.072}$ & $4.99_{-0.56}^{+0.65}\times10^{-1}$ & 1.69/3 \\
 & & & &
 &  7~GeV/$c$ & $1.060_{-0.124}^{+0.130}$ & $4.94_{-0.94}^{+1.21}\times10^{-1}$ & 3.07/3 \\
       &         &      &         & $\beta (dN_{ch}/d\eta/dN^{0}_{ch}/d\eta)^{\alpha}$ 
 &  5~GeV/$c$ & $0.940_{-0.035}^{+0.035}$ & $8.22_{-0.67}^{+0.74}\times10^{-1}$ & 15.46/3 \\
 & & & &
 &  6~GeV/$c$ & $0.917_{-0.061}^{+0.063}$ & $7.80_{-1.06}^{+1.26}\times10^{-1}$ & 2.26/3 \\
 & & & &
 &  7~GeV/$c$ & $0.931_{-0.108}^{+0.113}$ & $7.70_{-1.77}^{+2.39}\times10^{-1}$ & 3.81/3 \\
       &         &      &         & $\beta (\epsilon\tau_{0}/\epsilon^{0}\tau_{0})^{\alpha}$ 
 &  5~GeV/$c$ & $1.670_{-0.061}^{+0.063}$ & $9.83_{-0.86}^{+0.96}\times10^{-1}$ & 15.29/3 \\
 & & & &
 &  6~GeV/$c$ & $1.630_{-0.108}^{+0.112}$ & $9.28_{-1.35}^{+1.64}\times10^{-1}$ & 1.92/3 \\
 & & & &
 &  7~GeV/$c$ & $1.650_{-0.192}^{+0.202}$ & $9.18_{-2.25}^{+3.12}\times10^{-1}$ & 3.88/3 \\
\end{tabular} \end{ruledtabular}
\end{table*}

%--------------------------------------------  Table_VIII
\begin{table*}[!htb]
\caption{Parameters from fitting the indicated power-law functions 
for \dptpt to the data as a function of \ptpp for Pb$+$Pb collisions
at \sqsn=~2.76~TeV.}
\label{FitParTablexmodePb}
\begin{ruledtabular} \begin{tabular}{ccccccccccc}
System & \sqsn & year & hadron & \dptpt$=$ 
& \ptpp & $\alpha$ & $\beta$ & $\chi^{2}/ndf$ \\
\hline
Pb$+$Pb & 2.76~TeV & 2010-11  & $h^{+/-}$ & $\beta (\Npart/N^{0}_{\rm part})^{\alpha}$ 
 &  5~GeV/$c$ & $0.357_{-0.004}^{+0.004}$ & $2.44_{-0.04}^{+0.04}\times10^{-1}$ & 44.19/7 \\
 & & & &
 &  6~GeV/$c$ & $0.378_{-0.003}^{+0.004}$ & $2.74_{-0.04}^{+0.04}\times10^{-1}$ & 90.44/7 \\
 & & & &
 &  7~GeV/$c$ & $0.398_{-0.004}^{+0.004}$ & $2.96_{-0.04}^{+0.05}\times10^{-1}$ & 70.86/7 \\
 & & & &
 & 10~GeV/$c$ & $0.490_{-0.006}^{+0.006}$ & $3.16_{-0.04}^{+0.05}\times10^{-1}$ & 10.32/7 \\
 & & & &
 & 12~GeV/$c$ & $0.507_{-0.008}^{+0.008}$ & $3.04_{-0.04}^{+0.05}\times10^{-1}$ & 11.41/7 \\
 & & & &
 & 15~GeV/$c$ & $0.557_{-0.014}^{+0.014}$ & $2.82_{-0.04}^{+0.05}\times10^{-1}$ & 2.29/7 \\
       &         &      &         & $\beta (\Nqp/N^{0}_{\rm qp})^{\alpha}$ 
 &  5~GeV/$c$ & $0.320_{-0.003}^{+0.003}$ & $2.44_{-0.04}^{+0.04}\times10^{-1}$ & 34.51/7 \\
 & & & &
 &  6~GeV/$c$ & $0.339_{-0.003}^{+0.003}$ & $2.73_{-0.04}^{+0.04}\times10^{-1}$ & 71.06/7 \\
 & & & &
 &  7~GeV/$c$ & $0.358_{-0.003}^{+0.003}$ & $2.95_{-0.04}^{+0.05}\times10^{-1}$ & 59.14/7 \\
 & & & &
 & 10~GeV/$c$ & $0.440_{-0.005}^{+0.005}$ & $3.16_{-0.04}^{+0.05}\times10^{-1}$ & 9.62/7 \\
 & & & &
 & 12~GeV/$c$ & $0.456_{-0.007}^{+0.007}$ & $3.04_{-0.04}^{+0.05}\times10^{-1}$ & 13.94/7 \\
 & & & &
 & 15~GeV/$c$ & $0.501_{-0.013}^{+0.013}$ & $2.83_{-0.04}^{+0.05}\times10^{-1}$ & 2.30/7 \\
       &         &      &         & $\beta (dN_{ch}/d\eta/dN^{0}_{ch}/d\eta)^{\alpha}$ 
 &  5~GeV/$c$ & $0.298_{-0.003}^{+0.003}$ & $2.46_{-0.04}^{+0.04}\times10^{-1}$ & 66.71/7 \\
 & & & &
 &  6~GeV/$c$ & $0.313_{-0.003}^{+0.003}$ & $2.77_{-0.04}^{+0.04}\times10^{-1}$ & 145.00/7 \\
 & & & &
 &  7~GeV/$c$ & $0.329_{-0.003}^{+0.003}$ & $2.98_{-0.05}^{+0.05}\times10^{-1}$ & 123.28/7 \\
 & & & &
 & 10~GeV/$c$ & $0.404_{-0.005}^{+0.005}$ & $3.19_{-0.04}^{+0.05}\times10^{-1}$ & 30.94/7 \\
 & & & &
 & 12~GeV/$c$ & $0.417_{-0.006}^{+0.006}$ & $3.06_{-0.04}^{+0.05}\times10^{-1}$ & 26.21/7 \\
 & & & &
 & 15~GeV/$c$ & $0.455_{-0.011}^{+0.011}$ & $2.85_{-0.04}^{+0.05}\times10^{-1}$ & 5.76/7 \\
       &         &      &         & $\beta (\epsilon\tau_{0}/\epsilon^{0}\tau_{0})^{\alpha}$ 
 &  5~GeV/$c$ & $0.576_{-0.006}^{+0.006}$ & $2.43_{-0.04}^{+0.04}\times10^{-1}$ & 53.83/7 \\
 & & & &
 &  6~GeV/$c$ & $0.614_{-0.006}^{+0.006}$ & $2.73_{-0.04}^{+0.04}\times10^{-1}$ & 91.36/7 \\
 & & & &
 &  7~GeV/$c$ & $0.649_{-0.006}^{+0.006}$ & $2.96_{-0.04}^{+0.05}\times10^{-1}$ & 79.47/7 \\
 & & & &
 & 10~GeV/$c$ & $0.799_{-0.009}^{+0.009}$ & $3.17_{-0.04}^{+0.05}\times10^{-1}$ & 32.58/7 \\
 & & & &
 & 12~GeV/$c$ & $0.829_{-0.013}^{+0.013}$ & $3.05_{-0.04}^{+0.05}\times10^{-1}$ & 30.78/7 \\
 & & & &
 & 15~GeV/$c$ & $0.909_{-0.023}^{+0.023}$ & $2.83_{-0.04}^{+0.05}\times10^{-1}$ & 6.28/7 \\
\end{tabular} \end{ruledtabular}
\end{table*}

\clearpage

%%%%%%%%%%%%%%%%%%%%%%%%%%%%%%%%%%%% References

%\bibliography{ppg171x1}   

%merlin.mbs apsrev4-1.bst 2010-07-25 4.21a (PWD, AO, DPC) hacked
%Control: key (0)
%Control: author (0) dotless jnrlst
%Control: editor formatted (1) identically to author
%Control: production of article title (0) allowed
%Control: page (1) range
%Control: year (0) verbatim
%Control: production of eprint (0) enabled
%
 
\end{document}